\documentclass{ieeeaccess}
\usepackage{cite}
\usepackage{amsmath,amssymb,amsfonts}
\usepackage{algorithmic}
\usepackage{graphicx}
\usepackage{subcaption}
\usepackage{textcomp}
\usepackage{stfloats}
\usepackage{url}
\usepackage{verbatim}
\usepackage[detect-all]{siunitx}
\usepackage[hidelinks]{hyperref}
\usepackage{cleveref}
\usepackage{caption}

\DeclareCaptionFormat{figcapfont}{\fontsize{7}{8.4}\selectfont#1#2#3}
\DeclareCaptionFormat{subfigcapfont}{\fontsize{6}{7.2}\selectfont#1#2#3}
\def\BibTeX{{\rm B\kern-.05em{\sc i\kern-.025em b}\kern-.08em
    T\kern-.1667em\lower.7ex\hbox{E}\kern-.125emX}}
\usepackage{etoolbox}
\def\thevol{XX}
\def\myyear{2023}
\makeatletter
\patchcmd{\@evenfoot}{2016}{\myyear}{}{}
\patchcmd{\@oddfoot}{2016}{\myyear}{}{}
\makeatother

\usepackage{fancyhdr}
\fancypagestyle{firststyle}
{
 \fancyhf{} 
 \fancyfoot[LO]{\textcopyright 2023 IEEE. Personal use of this material is permitted. Permission from IEEE must be obtained for all other uses.}
 }

\begin{document}

\history{Date of publication xxxx 00, 0000, date of current version xxxx 00, 0000.}
\doi{10.1109/ACCESS.2017.DOI}

\title{On the Analysis and Optimization of Fast Conditional Handover with Hand Blockage for Mobility}
\author{\uppercase{Subhyal Bin Iqbal}\authorrefmark{1,2}, \IEEEmembership{Graduate Student Member, IEEE},
\uppercase{Salman Nadaf\authorrefmark{1}, Ahmad Awada \authorrefmark{1} \IEEEmembership{Member, IEEE}, Umur Karabulut \authorrefmark{1} \IEEEmembership{Member, IEEE}, Philipp Schulz \authorrefmark{2} \IEEEmembership{Member, IEEE} and Gerhard P. Fettweis \authorrefmark{2} \IEEEmembership{Fellow, IEEE}}.\authorrefmark{3}}
\address[1]{Nokia Standardization and Research Lab, 81541 Munich, Germany}
\address[2]{Vodafone Chair for Mobile Communications Systems, Technische Universität Dresden, 01062 Dresden , Germany}

\markboth
{S.B. Iqbal \headeretal: On the Analysis and Optimization of Fast Conditional Handover with Hand
Blockage for Mobility}
{S.B. Iqbal \headeretal: On the Analysis and Optimization of Fast Conditional Handover with Hand
Blockage for Mobility}

\corresp{Corresponding author: Subhyal Bin Iqbal (e-mail: subhyal.bin\_iqbal@nokia.com).}

\begin{abstract}
Although frequency range 2 (FR2) systems are an essential part of 5G-Advanced and future 3GPP releases, the mobility performance of multi-panel user equipment (MPUE) with hand blockage is still an area open for research and standardization. In this article, a comprehensive study on the mobility performance of MPUE with hand blockage is performed for conditional handover (CHO) and its potential enhancement denoted by fast conditional handover (FCHO). In contrast to CHO, in FCHO the MPUE can reuse earlier target cell preparations after each handover to autonomously execute subsequent handovers. This saves both the signaling overhead associated with the reconfiguration and re-preparation of target cells after each handover and reduces mobility failures. Results have shown that FCHO offers considerable mobility performance gains as compared to CHO for different hand blockage cases that are dependent on the hand position around the MPUE. For the worst-case hand blockage scenario, it is seen that mobility failures reduce by 10.5\% and 19.3\% for the 60 km/h and 120 km/h mobility scenarios, respectively. This gain comes at the expense of reserving the handover resources of an MPUE for a longer time given that the target cell configurations are not necessarily released after each handover. In this article, the longer resource reservation problem in FCHO is analysed and three different resource reservation optimization techniques are introduced. Results have shown that these optimization techniques not only reduce the resource reservation time but also significantly reduce the signaling overhead at the possible expense of a tolerable degradation in mobility performance.
\end{abstract}

\begin{keywords}
FR2, 5G-Advanced, mobility performance, multi-panel UE, hand blockage, fast conditional handover, signaling overhead, resource reservation optimization
\end{keywords}

\titlepgskip=-15pt

\maketitle
\thispagestyle{firststyle}

\section{Introduction}
\label{Sec1}
\PARstart{F}{requency}  range 2 (FR2) \cite{b1} addresses the problem of contiguous bandwidth that is required for 5G networks to fulfill the steep increase in user data throughput and low latency requirements. However, it also introduces additional challenges to the link budget design such as higher free-space path loss and penetration loss in mobile environments \cite{b2}. Another major challenge in FR2 is that at higher frequencies of the order of \SI{28}{GHz} the penetration depth into the human hand holding the user equipment (UE) is very small \cite{b3, b4}. This results in a high degree of blockage by the hand and significantly impairs the link margins in FR2.

Conditional handover (CHO) has been introduced in \cite{b5} as an alternative to baseline handover to improve the mobility performance of mobile systems in 5G networks. However, it introduces considerable overhead to the signaling that takes place both between the UE and the network and in between the network entities. This is due to the inherent decoupling of the handover preparation and execution procedures in CHO. Our earlier study \cite{b6} has covered fast conditional handover (FCHO), which is a potential enhancement to CHO for 5G-Advanced \cite {b7} whereby the UE maintains the configuration of the prepared target cells after a successful handover has taken place. This allows the UE to perform handovers consecutively and autonomously without requiring reconfiguration from the network. FCHO brings two advantages against CHO. Firstly, the preparation of multiple target cells and the signaling overhead involved in preparing multiple target cells is significantly reduced thanks to keeping the conditional configurations of the prepared cells after the handover. Secondly, the reuse of prepared target cells means that cells are now prepared relatively early and handovers can be executed immediately which otherwise in CHO would have resulted in a mobility failure. Consequently, mobility failures are also reduced by using FCHO. However, it has also been concluded in \cite{b6} that FCHO has the downside of excessive resource reservation time. This is because the UE maintains the prepared target cells for a longer time as compared to CHO and these target cells do not release the resources of those preparations until either the CHO release or replace conditions \cite{b6} are fulfilled.

On the other hand, 3GPP has proposed a stochastic hand blockage model in \cite{b8} that captures the spatial region of the blockage around the UE in a local coordinate system for both the portrait and landscape modes, where a \SI{30}{dB} flat loss is assumed over the spatial region. The studies in \cite{b3, b4, b9, b10, b11} have focused extensively on the modeling and remedy of hand blockage for UEs with form factor considerations in FR2. The hand blockage models used have been either electromagnetic (EM) simulation-based hand blockage models or measurement-based hand blockage models. However, in none of these studies a system-level mobility performance analysis of UEs with hand blockages has been carried out in a 5G-Advanced network. Such studies are critical to both understand and address the problems imposed by hand blockage on mobility as a result of impaired link margins. Although a mobility performance analysis of CHO has been performed earlier in \cite{b6, b12, b13, b14}, the studies listed did not consider hand blockage. To our knowledge, the only study to date on the mobility performance of FCHO has been our earlier work in \cite{b6}, wherein the hand blockage effect was still not considered. In our first contribution to this article, we study the impact of hand blockage on the mobility performance and analyse the benefit of FCHO over CHO in terms of enhancing the mobility performance for different hand blockage cases. Herein, the Cellular Telecommunication Industry Association (CTIA) defined wide-grip hand phantom \cite{b15} model is used and simulation results are generated for different hand positions and user speeds in a 5G network. The hand phantom model is modeled using \textit{CST Studio Suite} (a commercial-grade electromagnetic simulation software suite) \cite{b16}. Multi-panel UEs  \cite{b17, b18} are an essential part of 5G-Advanced and future 3GPP releases and therefore this paper considers an MPUE with three antenna panels in an \textit{edge} design. A detailed analysis of CHO and FCHO is carried out for an extended set of mobility key performance indicators (KPIs) and the key benefits of FCHO are highlighted in terms of combating hand blockage.

For the efficient distribution of resources in the network, it is important that resource reservation time is optimized. It is known from \cite{b12, b14, b19} that the resource reservation time in CHO is high since multiple cells reserve resources for a single UE. In \cite{b6} it was concluded that in FCHO the resource reservation problem is exacerbated due to the retention of prepared cells after a handover. Moreover, it was said further studies may be needed to address this problem. This article is a first step in this direction. To the best of our knowledge, FCHO resource reservation optimization has not been studied in the current literature. In our second contribution of this article, the resource reservation problem in FCHO that was earlier identified in \cite{b6} is studied in detail and three different approaches are introduced which offer a tradeoff between the mobility performance, signaling overhead, and resource reservation time. These optimization approaches adopt the principles of mobility robustness optimization (MRO) \cite{b20, b21} algorithms whereby cell preparations may be blocked, selectively discarded after a handover, or delayed based on statistics that have been collected from past mobility events. 

This article is structured as follows. In Section \ref{Sec2}, hand blockage in FR2 is explained and the most commonly used hand blockage models in the current literature are presented. Furthermore, the differences between CHO and FCHO are explained in terms of the key CHO signaling events. In Section \ref{Sec3}, the system model for the 5G network is introduced and the different hand blockage cases are discussed. In Section \ref{Sec4}, the KPIs are explained and the mobility performance of FCHO is compared with CHO for different hand blockage use cases. In Section \ref{Sec5}, the resource reservation problem in FCHO is described. Then, the three different FCHO resource reservation optimization approaches are explained. Thereafter, a performance analysis of the FCHO resource reservation optimization approaches is provided in Section \ref{Sec6}, and the results are discussed taking into account the tradeoff between mobility performance, signaling overhead, and resource reservation. Finally, the paper is concluded in Section \ref{Sec7}.

\section{Background and Motivation}
\label{Sec2}

In this section, the impact of hand blockage in FR2 is explained and the main hand blockage models used in current literature are discussed. Further on, the key differences between CHO and FCHO are highlighted in terms of CHO preparation, release, and replace events.

\subsection{Hand Blockage Modeling in FR2} \label{Subsec2.1}

FR2 systems have significantly matured in the last few years and the first wave of commercial deployments are currently available in the market across multiple geographical locations. However, a number of basic issues in terms of their practical viability remains a topic of ongoing research. One such issue is that of hand blockages, which as discussed in \Cref{Sec1}, can significantly impair link margins in FR2 compared to lower carrier frequencies. This is because the relative skin permittivity of the human hand decreases with an increase in the frequency of the radio waves, meaning that at FR2 carrier frequencies of the order of \SI{28}{GHz} the blockage effect of the human hand is much greater than that of lower frequencies \cite{b22}. Hence, for any mobility studies based on MPUE in FR2, it is imperative to have a suitable hand blockage model that captures the spatial region that is lost due to blockage and the associated loss in the reference signal received power (RSRP) over this region. If the current literature is taken into account, hand blockage models can be broadly categorized into three categories:
\begin{itemize}
    \item stochastic hand blockage models \cite{b8},
    \item electromagnetic simulation-based hand blockage models \cite{b4, b9},
    \item measurement-based hand blockage models \cite{b10, b11}.
    
\end{itemize}

Each of these hand blockage models is explained in detail below.

\subsubsection{Stochastic Hand Blockage Models} 

\begin{figure*}[!b]    \begin{subfigure}{0.5\textwidth}        \centering        \includegraphics[width=0.8\textwidth]{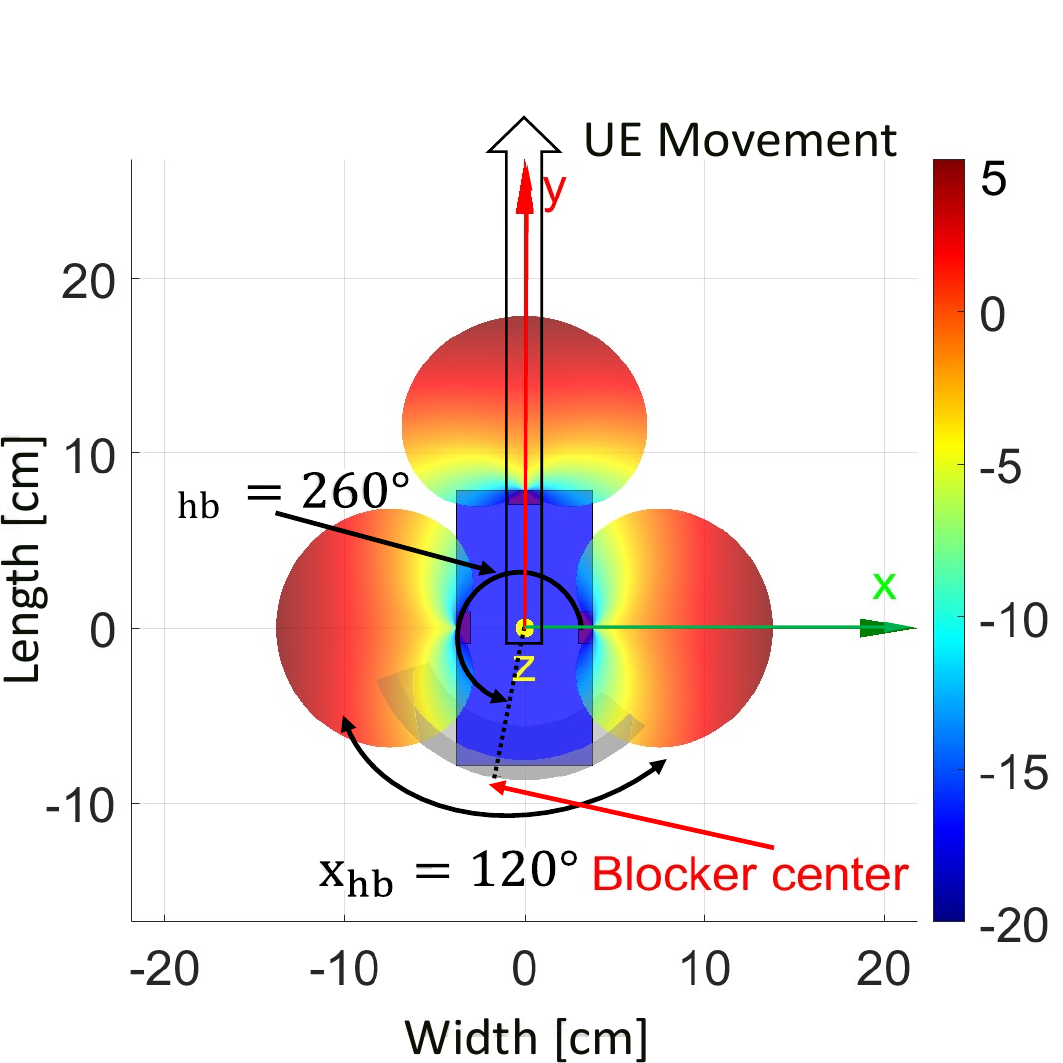}           \caption{\textit{Top} view.}       \label{fig:Fig0a}         \end{subfigure}\hfill    
\begin{subfigure}{0.5\textwidth}        
\centering        \includegraphics[width=0.8\textwidth]{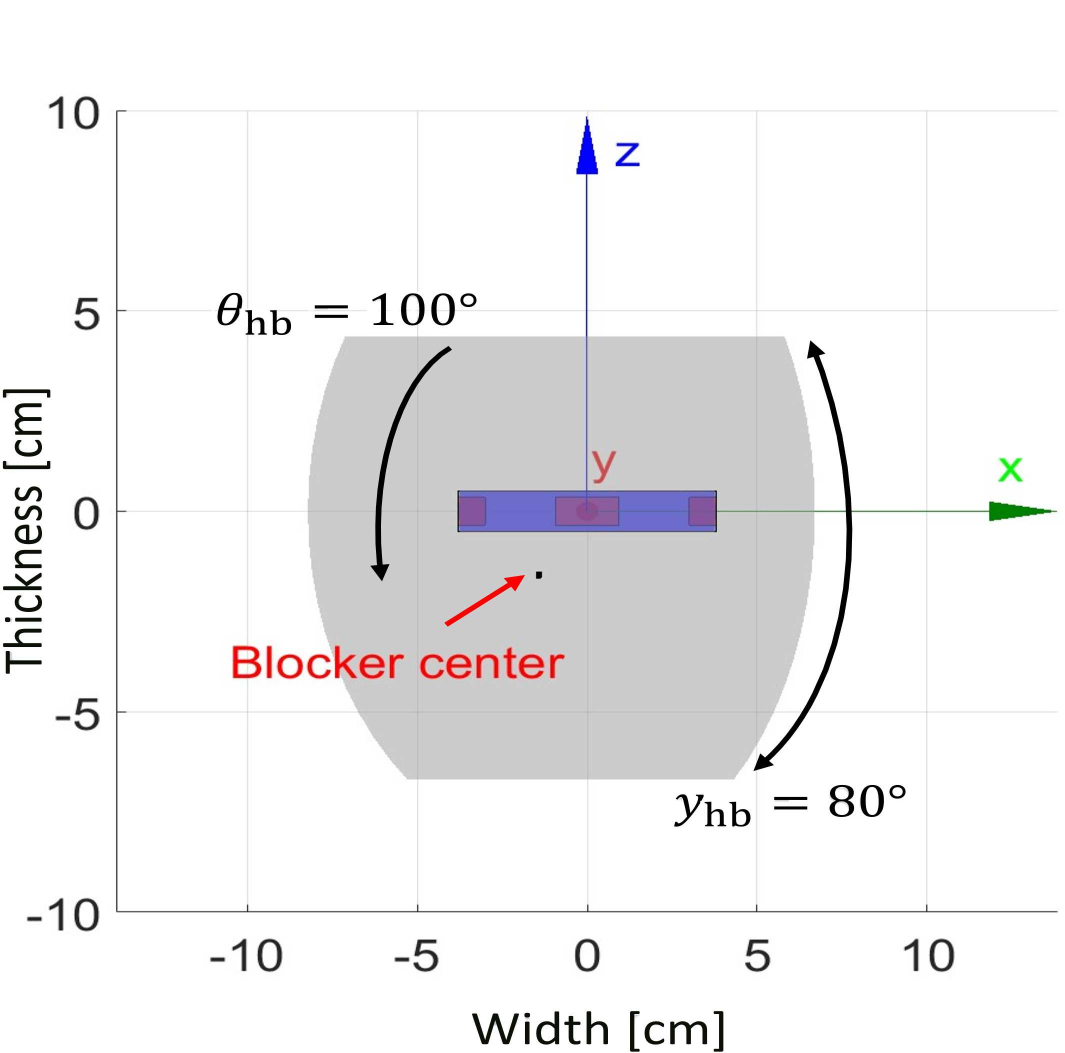}             \caption{\textit{Side view}.}         \label{fig:Fig0b}     \end{subfigure}\hfill 
\caption{MPUE with edge design \cite{b17, b18} in the portrait orientation (parallel to the ground) along with the spherical blockage region defined by 3GPP shown as (a) top view and (b) side view. $\phi_{\mathrm{hb}}$ is the hand blocker orientation and $x_{\mathrm{hb}}$ is the angular span of the blocker in azimuth. $\theta_{\mathrm{hb}}$ is the hand blocker orientation and $y_{\mathrm{hb}}$ is the angular span of the blocker in elevation.
}   \label{fig:Fig0}    \end{figure*}

The most well-known stochastic hand blockage model is the stochastic variant of the 3GPP hand blockage model \cite[pp. 62-64]{b8} for 5G networks. It proposes a spherical blockage that is tailored to the human hand in portrait and landscape orientations around a UE. A \SI{30}{dB} abrupt flat loss in the RSRP is assumed if the angle of arrival intersects this spherical blockage region. Fig.\,\ref{fig:Fig0} shows the blockage region for the 3GPP hand blockage model defined for an MPUE with three panels in \textit{edge} design, where the MPUE in portrait orientation has three directional antenna panels on its left, top, and right edges.

The 3GPP hand blockage model is pessimistic with the 30~dB abrupt flat loss that it assumes over the spherical region because it mostly considers studies that take horn antennas into account \cite{b23, b24} that are used to generate the blockage model. It has been shown in more recent studies \cite{b4} that the blockage loss associated with horn antennas is substantially less than \SI{30}{dB}. Secondly, the abruptness of the model is not well-suited for mobility since in real-world mobile environments the RSRP degradation due to hand blockage is relatively smooth \cite{b3}. In mobility studies, this would then lead to a pessimistic evaluation of the mobility failures, particularly for high UE speeds. This is discussed later in this section.

\subsubsection{EM Simulation-Based Hand Blockage Models}

EM simulation-based hand blockage models are based on simulation studies that consider different hand grips modeled around the UEs with form factor considerations taken into account in commercial-grade EM software simulators such as \textit{CST Studio Suite} \cite{b16}. These simulators model the human hand based on the hand phantom (how the UE is gripped) and dielectric properties of the skin tissue, e.g., relative dielectric constant at different frequencies as well as dielectric properties of UE materials including the antenna panels. The hand dielectric properties determine the penetration depth into the hand and the reflection of electromagnetic waves from the hand. Thereafter, the antenna element radiation patterns for each individual element of the panels can be determined and included in the link budget design. Different hand phantom models are defined by CTIA \cite{b15} based on the UE width and usage, i.e., talk or data mode. Since most modern UEs have a width between \SIrange{7.3}{9.2}{cm}, the wide-grip \cite[pp. 364]{b15} hand phantom model is used most commonly in current literature and this article also considers a wide-grip where it is assumed that the hand grip remains the same for talk or data mode.

\begin{figure}[!t]    
\begin{subfigure}{0.5\textwidth}        
\centering        
\includegraphics[width=1\textwidth]{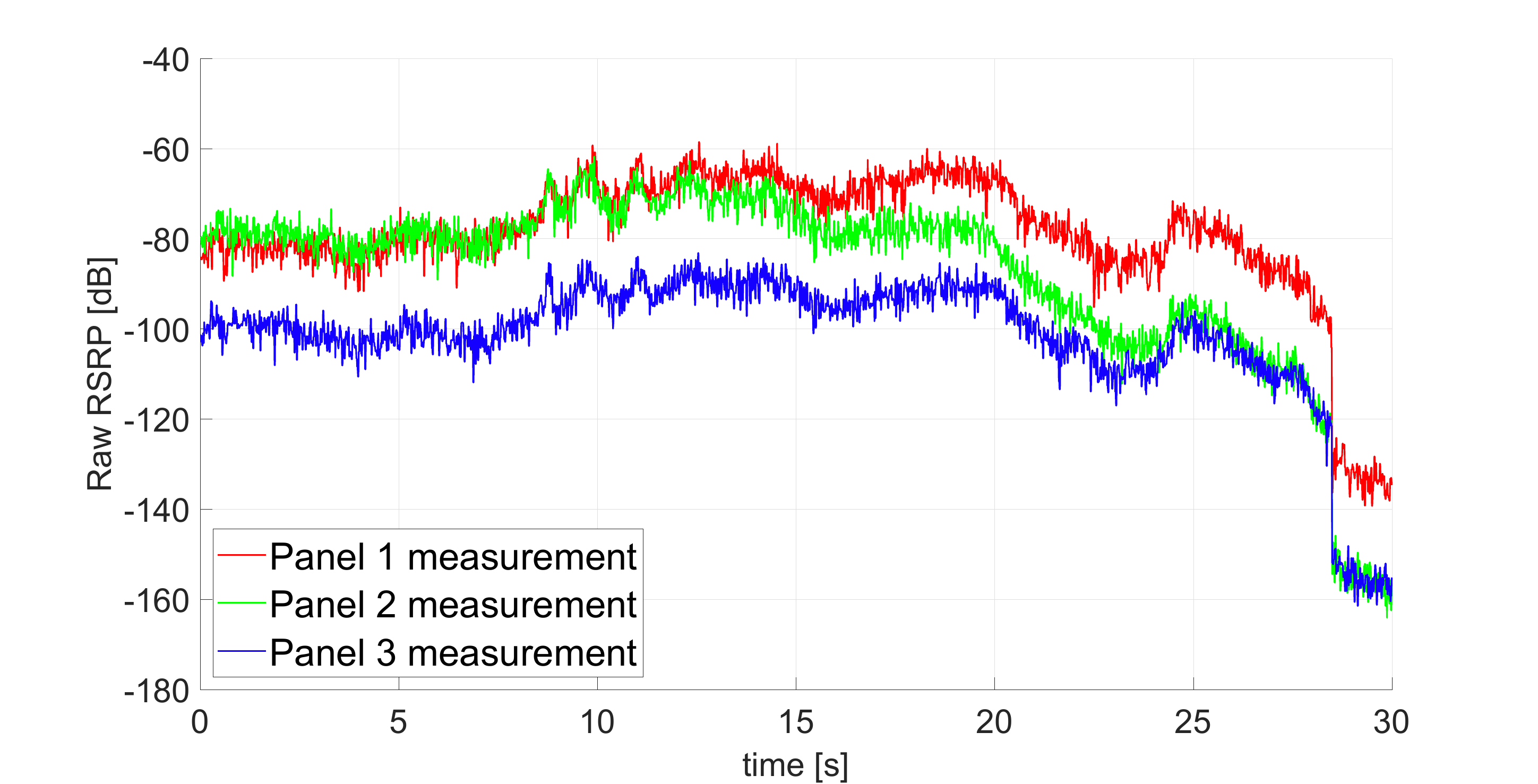}        \label{fig:Fig2a}
\vspace{-8pt}
\caption{3GPP hand blockage model.} 
\label{fig:Fig2a}    
\end{subfigure}\hfill    

\begin{subfigure}{0.5\textwidth}        
\centering        
\includegraphics[width=1\textwidth]{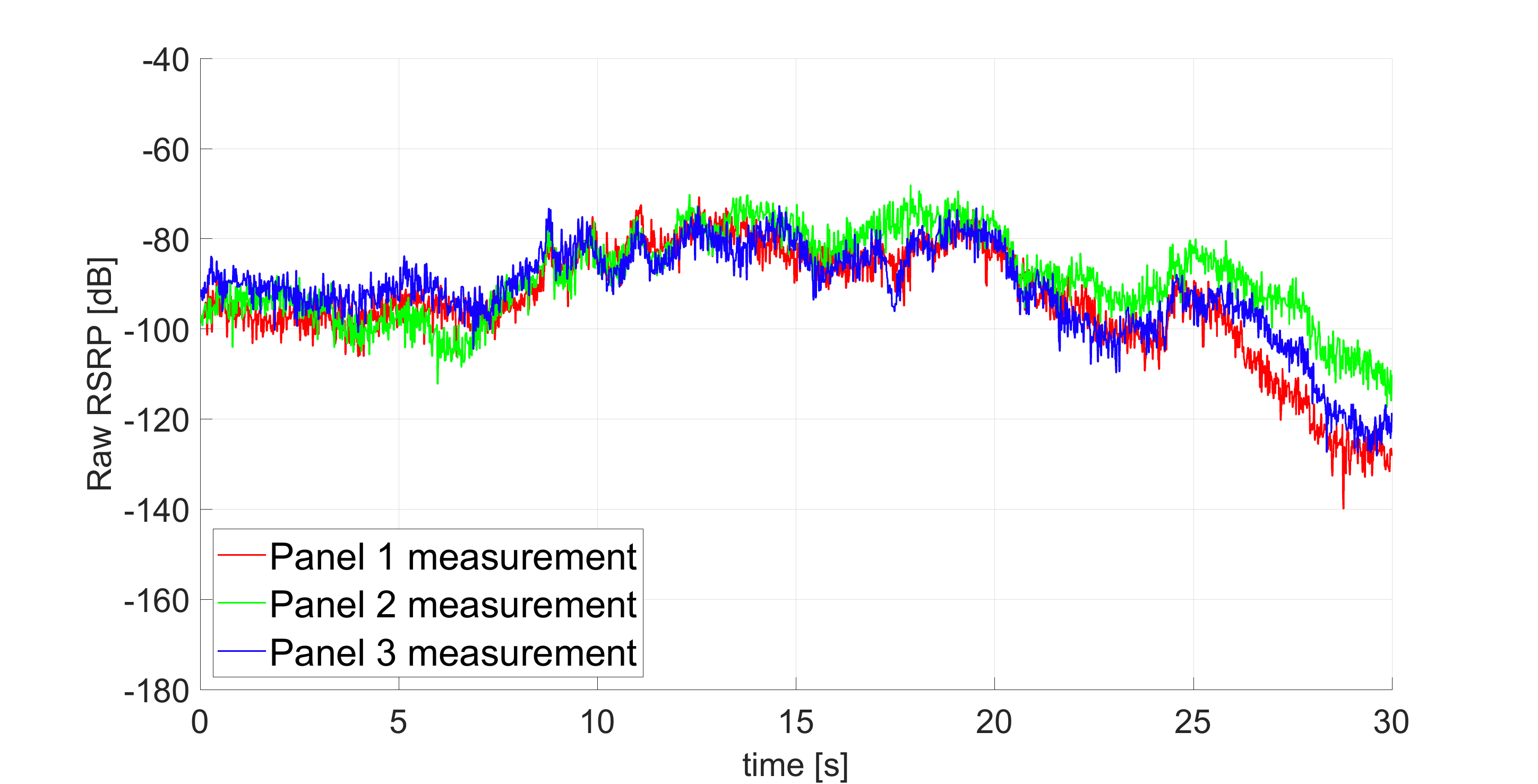}        \label{fig:Fig2b}        
\vspace{-8pt}
\caption{EM simulation-based hand blockage model.} 
\label{fig:Fig2b}    
\end{subfigure}\hfill

\caption{Raw RSRP degradation for (a) 3GPP hand blockage model and (b)  EM simulation-based hand blockage model for 30 km/h mobility use case.} 
 \label{fig:Fig2}    
\vspace{-10pt}
 \end{figure}

In Fig.\,\ref{fig:Fig2} the measured raw RSRPs (unfiltered physical layer measurements) of an MPUE for three different panels over time is depicted for the serving cell {$c_0$}. Fig.\,\ref{fig:Fig2a} shows the RSRP degradation with the 3GPP hand blockage model, and Fig.\,\ref{fig:Fig2b} shows the RSRP degradation with an EM simulation-based hand blockage model. Both blockage models consider a case where all three MPUE panels experience blockage for a \SI{30}{km/h} mobility use case. It can clearly be seen that for the 3GPP hand blockage model in Fig.\,\ref{fig:Fig2a}, the RSRP degrades abruptly over a few ms by approximately \SI{30}{dB} whereas roughly the same degradation (in the range from 27-\SI{36}{dB}) for the EM hand blockage model occurs over a period of approximately \SI{5}{s}. Consequently, a mobility failure is experienced in Fig.\,\ref{fig:Fig2a} whereas the UE avoids a mobility failure in Fig.\,\ref{fig:Fig2b} by performing a handover to another cell. 

\subsubsection{Measurement-Based Hand Blockage Models}

Measurement-based blockage models consider commercial-grade UEs where measurements are performed in an anechoic chamber for FR2 deployments. In such models, the electric field information (amplitude and phases) that has been measured inside the anechoic chamber is then used to generate the antenna element radiation patterns for each panel of an MPUE. The study in \cite{b10} considers an anthropomorphic hand phantom composed of a silicone-carbon-based mixture with material properties conforming to CTIA definitions and standards. This hand phantom grips an MPUE in \textit{edge} design with three panels with the help of a robotic positioning mechanism. The study in \cite{b11} considers a measurement setup in an anechoic chamber where a test human subject holds an MPUE in various CTIA-defined grips. In this article, we do not include an evaluation with measurement-based hand blockage models since the EM simulation-based hand blockage model that we have defined for different hand grips is already well suited to our simulation environment where we consider multiple mobile UEs in a 5G network.

\subsection{CHO and FCHO} \label{Subsec2.2}

In CHO, multiple target cells can be prepared for a potential handover to one of these prepared target cells \cite{b5, b6, b12}. Three distinct signaling events can be defined for cell preparation. In each of the signaling events, the layer 3 (L3) cell quality RSRP $P_{c}^\textrm{L3}(m)$ is used \cite{b6}, where cell $c \in C$. Here $m$ is defined as $m=n\omega$, where $n$ is the discrete-time instant at which the UE measures the raw RSRP measurements transmitted over the synchronization signal block (SSB) bursts and $\omega$ is the L1 measurement period (aligned with the SSB periodicity). The CHO signaling events are discussed below. 

\begin{itemize}

\item
\textit{CHO Preparation Event}

The CHO \textit{preparation} event is required so that the UE being served by the serving cell $c_0$ can initiate the preparation of the target cell $c^{\prime}$ for handover. The CHO preparation condition that is monitored by the UE is defined as

\begin{multline}
\label{Eq1}
     P_{c_0}^\textrm{L3}(m)  < P_{c^{\prime}}^\textrm{L3}(m) + o^\mathrm{prep}_{c_0,c^{\prime}} \\ \text{for} \  m_\textrm{prep} - T_\mathrm{prep} < m < m_\textrm{prep},
\end{multline} 

where $o^\mathrm{prep}_{c_0,c^{\prime}}$ is the CHO preparation offset between cell $c_0$ and $c^{\prime}$. The UE sends a measurement report to the serving cell $c_0$ at time $m=m_\mathrm{prep}$ if the preparation condition is fulfilled for the preparation condition monitoring time $T_\mathrm{prep}$. Having received the measurement report, the serving cell initiates the preparation of target cell $c^{\prime}$ over the Xn interface and provides the UE with the conditional configuration of target cell $c^{\prime}$. 

\item
\textit{CHO Release Event}

In case the RSRP of any prepared target cell $c^{\prime}$ degrades after preparation, the resources that are allocated for handover by cell $c^{\prime}$ for that particular UE should be released so that they can be reused by other UEs in the network. This ensures efficiency in resource usage. The CHO \textit{release} event is triggered by the CHO release condition that is defined as

\begin{equation}
\label{Eq2}
     P_{c^{\prime}}^\textrm{L3}(m) + o^\mathrm{rel}_{c_0,c^{\prime}}   < P_{c_0}^\textrm{L3}(m) \ \text{for} \ m_\textrm{rel} - T_\mathrm{rel} < m < m_\textrm{rel},
\end{equation}
where $o^\mathrm{rel}_{c_0,c^{\prime}}$ is the CHO release offset between cell $c_0$ and $c^{\prime}$. The UE sends a measurement to the serving cell $c_0$ at  $m=m_\mathrm{rel}$ if the release condition is fulfilled for the release condition monitoring time $T_\mathrm{rel}$. Having received the measurement report, the serving cell cancels the preparation at the target cell $c^{\prime}$ and reconfigures the UE to release the configuration of target cell $c^{\prime}$.

\item
\textit{CHO Replace Event}

The number of prepared target cells that can be prepared for CHO is limited to restrict the resource reservations for the same UE in the network \cite{b12}. It is essential that the list of prepared cells is kept up-to-date, even when the maximum number of cell preparations is reached in order to minimize mobility failures in the network. Therefore, the weakest prepared cell $c_\textrm{W}$ in the list can be replaced by another stronger neighboring cell $c_\textrm{S}$ through the CHO replace condition which is modeled as

\begin{equation}
\label{Eq3}
     P_{c_\textrm{S}}^\textrm{L3}(m) > P_{c_\textrm{W}}^\textrm{L3}(m) + o^\mathrm{rep}_{c_\textrm{W,S}} \ \text{for} \ m_\textrm{rep} - T_\mathrm{rep} < m < m_\textrm{rep},
\end{equation}
where $o^\mathrm{rep}_{c_\textrm{W,S}}$ is the CHO replace offset between cell $c_\textrm{W}$ and $c_\textrm{S}$. The UE sends a measurement report to the serving cell $c_0$ at time $m=m_\mathrm{rep}$ if the replace condition is fulfilled for the replace condition monitoring time $T_\mathrm{rep}$. Having received the measurement report, the serving cell can initiate the preparation of the new target cell $c_\textrm{S}$ and the release of the preparation in the old target cell $c_\textrm{W}$. The UE is then reconfigured by the serving cell $c_0$ to replace the conditional configuration of cell $c_\textrm{W}$ with that of $c_\textrm{S}$.

\end{itemize}

A more detailed explanation of the CHO signaling events along with their respective diagrams can be found in \cite[Section II]{b6}.

\begin{figure}[!b]
\textit{\centering
\includegraphics[width = 0.96\columnwidth]{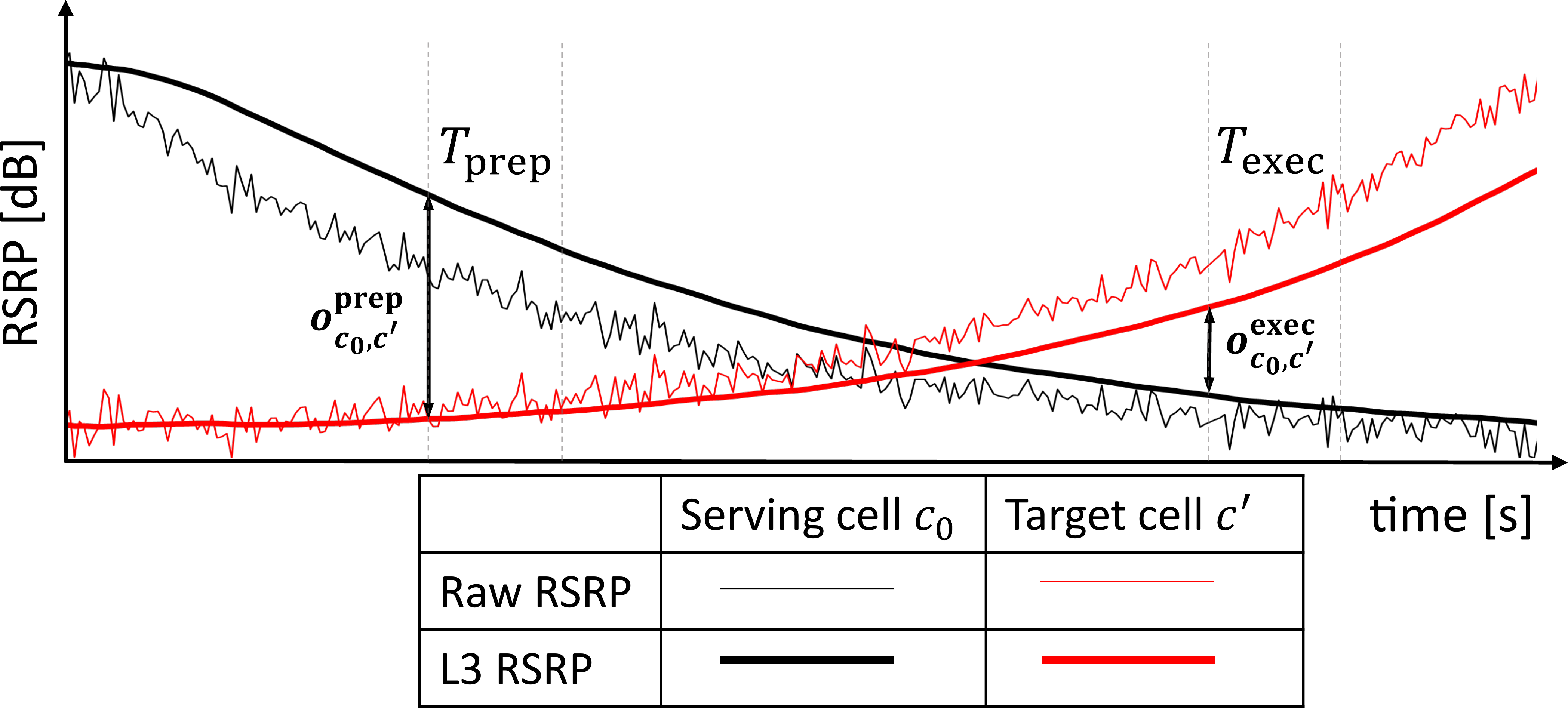}
\caption{Illustration of CHO process from serving cell $c_0$ to target cell $c^{\prime}$, where it can be seen that the handover preparation and execution phases are decoupled.} 
\label{fig:Fig2.1}} \vspace{-12pt}
\end{figure}

As mentioned in \Cref{Sec1}, CHO handover execution is decoupled from CHO preparation whereby the handover is prepared early but the actual handover execution takes place only when the radio link is sufficient. The UE monitors the CHO execution condition, defined as

\begin{equation}
\label{Eq4}
     P_{c_0}^\textrm{L3}(m) + o^\mathrm{exec}_{c_0,c^{\prime}}  < P_{c^{\prime}}^\textrm{L3}(m) \ \text{for} \ m_\textrm{exec} - T_\mathrm{exec} < m < m_\textrm{exec},
\end{equation}
where $o^\mathrm{exec}_{c_0, c^{\prime}}$ is defined as the CHO execution offset between cell $c_0$ and $c^{\prime}$. The UE executes a handover towards the prepared target cell $c^{\prime}$ if the execution condition at $m=m_\mathrm{exec}$ is fulfilled for the execution condition monitoring time $T_\mathrm{exec}$. The decoupled CHO process is shown in Fig.\,\ref{fig:Fig2.1}. 

FCHO has been defined in \cite{b25, b26} as a handover mechanism where it “might be possible to keep CHO candidates after the handover” and reuse target cell preparations instead of releasing them after a handover. In FCHO, the preparation of the target cell as well as the previous serving cell is maintained after every successful handover. When compared to conventional CHO, FCHO can be stated to have two advantages. Firstly, FCHO reduces mobility failures in cell border regions where high and rapidly increasing inter-cell interference may impact the ability of the network to successfully receive the measurement report from the UE or to provide a handover command timely to the UE. By keeping the target cell preparations as well as the preparation of the previous serving cell after the handover, FCHO allows the UE to perform a subsequent cell change immediately without waiting for being reconfigured by the network. It is useful to retain the previous serving cell configuration after a handover since it could be the next target cell. In conventional CHO it would mean first preparing one or more of such neighboring cells as a target cell, leading to a late preparation that could potentially result in a mobility failure. Secondly, FCHO reduces the overhead that is caused by CHO signaling events because multiple target cells need not be prepared after every handover thanks to their retention after a successful handover. This is beneficial since in conventional CHO it is highly likely that a prepared target cell before a handover (or the previous serving cell) will be re-prepared on account of fulfilling the CHO preparation condition given in (\ref{Eq1}).  

Compared to conventional CHO, the CHO preparation events are reduced in FCHO because the list of prepared cells is not released after every successful handover. From the UE perspective, the list of prepared cells for UE $u$ at time $m$ can be defined as 

\begin{equation}
\label{Eq4.5}
  n_u^\mathrm{prep}(m) \subseteq \{ 1, \dots, N_\mathrm{cells} \}
  \quad\text{with}\quad
  |n_u^\mathrm{prep}(m)| \leq n_u^{\mathrm{max}},
\end{equation}
where $N_\mathrm{cells}$ is the total number of UEs in the network.

Similarly, from the perspective of cell $c$ a list of UEs for which resources can be reserved can be defined as

\begin{equation}
\label{Eq4.6}
  n_c^\mathrm{prep}(m) \subseteq \{ 1, \dots, N_\mathrm{UE} \}
  \quad\text{with}\quad
  |n_c^\mathrm{prep}(m)| \leq n_c^{\mathrm{max}},
\end{equation}
where $N_\mathrm{UE}$ is the total number of UEs in the network. 

As the maximum number of prepared cells on the UE side $n_u^{\mathrm{max}}$ can be up to eight cells as defined by 3GPP \cite{b1}, this would mean that up to eight separate CHO preparation events can be avoided after each successful handover. Less CHO preparations also mean fewer CHO removals. On the other hand, the radio link of some of the retained cells may degrade due to UE movement or changing radio link conditions and they may need to be released. Similarly, there may be more CHO replace events in FCHO because some prepared target cells become weak over time. However, it is known from \cite{b12} that the number of CHO preparation events is much higher than that of CHO release and CHO replace events and they account for most of the signaling overhead. Therefore, the overall signaling overhead in FCHO will be much less than in conventional CHO. However, this comes at the expense of an increase in the resource reservation time. A diagrammatic explanation of FCHO signaling along with a detailed performance analysis of the signaling overhead for CHO and FCHO for two different mobility scenarios can be found in \cite[Section III]{b6} and \cite[Section V]{b6}, respectively.

\section{System Model} 
\label{Sec3}

In this section, the simulation setup for the 5G network model is explained along with the simulation parameters that are used later in the performance analysis. Thereafter, the different hand blockage cases are discussed.

\subsection{5G Network Model}
\label{Subsec3.1}

\begin{figure}[!t]
\textit{\centering
\includegraphics[width = 0.96\columnwidth]{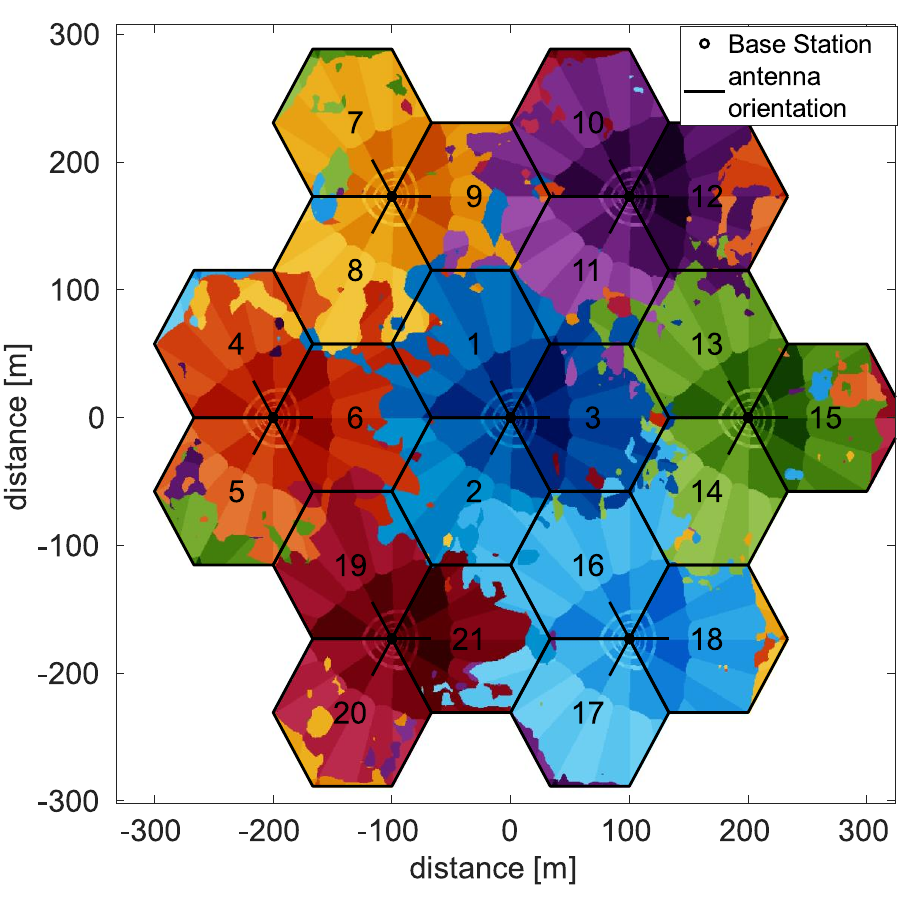}
\vspace{-0.4\baselineskip}
\caption{Simulation scenario consisting of seven hexagonal sites, where each site is serving 3 cells with 120$^{\circ}$ coverage. Tx-side beamforming is considered, consisting of 12 beams in each cell.} 
\label{fig:Fig3}} \vspace{-12pt}
\end{figure}

We consider a 5G network model with an urban-micro (UMi) cellular deployment consisting of a standard hexagonal grid with seven base station (BS) sites, each divided into three sectors or cells. The inter-cell distance is 200 meters and the FR2 carrier frequency is \SI{28}{GHz}. 420 UEs are dropped randomly following a 2D uniform distribution over the network at the beginning of the simulation, moving at constant velocities along straight lines where the direction is selected randomly at the start of the simulation \cite[Table 7.8-5]{b8}. A wrap-around \cite[pp. 140]{b265} is considered, i.e., the hexagonal grid with seven BS sites is repeated around the original hexagonal grid shown in Fig.\,\ref{fig:Fig3} in the form of six replicas. This implies that the cells on network borders are subject to interference from the other edge of the network that is comparable to the cells not on the network borders. From a simulation modeling perspective, if a UE moves out of the network border, it enters back from the other edge of the network. It is known from \cite{b6} that the rural and suburban scenarios are not very demanding in terms of the low interference regime because typically the number of simultaneously scheduled beams per cell is taken as $K_b=$ 1. As a result, CHO by itself addresses many of the mobility failures in the network. Therefore, two demanding yet realistic mobility scenarios with UE speeds based on \cite{b305} are considered. UEs moving at \SI{60}{km/h} represent the urban mobility scenario, which is the usual speed in the non-residential urban areas of cities. Whereas UEs moving at \SI{120}{km/h} represent the highway mobility scenario, which is the usual speed limit on major highways. The number of simultaneously scheduled beams per cell is taken as $K_b=$ 4. The simulation parameters are summarized in \Cref{Table1}.

\begin{table}[!t]
\vspace{-0.8\baselineskip}
\renewcommand{\arraystretch}{1.2}
\caption{Simulation parameters}
\vspace{-0.6\baselineskip}
\centering
\begin{tabular}{| l | l |}
\hline
\bfseries Parameter & \bfseries Value\\
\hline\hline
Carrier frequency & \SI{28}{GHz}\\
\hline
System bandwidth & \SI{100}{MHz}\\
\hline
Cell deployment topology & 7-site hexagon\\
\hline
Total number of cells $N_\mathrm{cells}$ & 21 \\
\hline
Downlink Tx power & \SI{40}{dBm}\\
\hline
Tx (BS) antenna height & 10~m\\
\hline
Tx antenna element pattern & Table 7.3-1 in \cite{b8} \\
\hline
Tx panel size & 16 $\times$ 8, $\forall b \in \{1,\ldots,8\}$\\
 & 8 $\times$ 4, $\forall b \in \{9,\ldots,12\}$ \\
\hline
Tx antenna element spacing & vertical: 0.7$\lambda$\\ 
& horizontal: 0.5$\lambda$\\
\hline
Beam elevation angle $\theta_b$ & 90$^{\circ}$, $\forall b \in \{1,\ldots,8\}$ \\
 & 97$^{\circ}$, $\forall b \in \{9,\ldots,12\}$\\
\hline
Beam azimuth angle $\phi_b$ & $-$52.5$^{\circ}$$+$15$(b-1)^{\circ}, \forall b \in \{1,\ldots,8\}$\\
& $-$45$^{\circ}$$+$30$(b-9)^{\circ}, \forall b \in \{9,\ldots,12\}$\\
\hline
Tx-side beamforming  & Fitting model of \cite{b27}\\
gain model & \\
\hline
Rx (UE) antenna height & \SI{1.5}{m}\\
\hline
Rx antenna element pattern & MPUE: based on \cite{b30} \\
\hline
Rx panel size  & Single antenna element\\
\hline
Rx antenna element gain  & MPUE: \SI{5}dBi\\ 
\hline
Total number of UEs $N_\mathrm{UE}$ & 420\\
\hline
UE speed  & urban scenario: \SI{60}{km/h}\\
& highway scenario: \SI{120}{km/h}\\
\hline
Number of simultaneously   & 4\\
scheduled beams per cell $K_b$ & \\ 
\hline
CHO preparation offset $o^\mathrm{prep}_{c_0,c^{\prime}}$  & \SI{10}{dB}\\
\hline
CHO execution offset $o^\mathrm{exec}_{c_0,c^{\prime}}$  & \SI{3}{dB}\\
\hline
CHO release offset $o^\mathrm{rel}_{c_0,c^{\prime}}$  & \SI{13}{dB}\\
\hline
CHO replace offset $o^\mathrm{rep}_{c_\textrm{W,S}}$  & \SI{3}{dB}\\
\hline
CHO preparation time $T_\mathrm{prep}$ & \SI{80}{ms}\\
\hline
CHO execution time $T_\mathrm{exec}$ & \SI{80}{ms}\\
\hline
CHO release time $T_\mathrm{rel}$ & 80~ms\\
\hline
CHO replace time $T_\mathrm{rep}$ & \SI{80}{ms}\\
\hline
Maximum prepared cells & 4\\
per UE $n_u^{\mathrm{max}}$ & \\ 
\hline
Fast-fading channel model & Abstract model of \cite{b27}\\
\hline
Time step $\Delta t$  & \SI{10}{ms} \\
\hline
SSB periodicity  & \SI{20}{ms} \\

\hline
Simulated time $t_\mathrm{sim}$ & \SI{30}{s} \\
\hline
SINR threshold $\gamma_\mathrm{out}$  & $-$\SI{8}{dB} \\
\hline
\end{tabular}
\label{Table1}
\end{table}

As per 3GPP's study outlined in \textit{Release 15} \cite{b8}, the  channel model we consider in this article takes into account shadow fading due to large obstacles (including the human body) and assumes a soft line-of-sight (LoS) for all radio links between the cells and UEs. Soft LoS is a weighted average of the LoS and non-LoS channel components \cite[pp. 59-60]{b8} and is used for both shadow fading and distance-dependent path loss calculation in our simulation scenario. Fast fading is taken into account through the low complexity channel model for multi-beam systems proposed in \cite{b27}, which integrates the spatial and temporal characteristics of 3GPP's geometry-based stochastic channel model (GSCM) \cite{b8} into Jake’s channel model \cite{b28}. The transmitter (Tx)-side beamforming gain model is based on the study conducted in \cite{b29}, where a 12-beam grid configuration is considered. Each beam $b \in B$ for cell $c \in C$. Beams $b \in \{1,\ldots,8\}$ have smaller beamwidth and higher beamforming gain and cover regions further apart from the BS. Beams  $b \in \{9,\ldots,12\}$ have larger beamwidth and relatively smaller beamforming gain and cover regions closer to the BS. This can also be seen in Fig.\,\ref{fig:Fig3}, where the eight outer beams are shown in light colors and the four inner beams are shown in dark colors. The effect of shadow fading is also visible as coverage islands in Fig.\,\ref{fig:Fig3}.

On the UE-side, the MPUE architecture is considered which assumes an \textit{edge} design with three directional antenna panels, each with a single antenna element and a maximum gain of \SI{5}{dBi} \cite{b17, b18}. The antenna element radiation pattern for each of the three panels is based on \cite{b30}. The UE screen, held by the user, is assumed to be parallel to the ground \cite{b17}. In line with 3GPP \cite{b306}, the signal measurement scheme considered is MPUE-A3, where the UE can measure the RSRPs from the serving cell $c_0$ and neighboring cells on all three panels simultaneously. A more detailed explanation of the MPUE-A3 signal measurement scheme can be found in \cite[pp. 3-4]{b17}. 

The average downlink signal-to-interference-plus-noise ratio (SINR) $\gamma_{c,b}(m)$ of a link between the UE and beam $b$ of cell $c$ is evaluated by the Monte-Carlo approximation given in \cite{b29} for a resource scheduler where all UEs get precisely the same amount of resources. This SINR is of key importance in the two mobility failures modeled on the handover failure (HOF) and radio link failure (RLF) models in this article, each of which is elaborated below.

\textit{HOF Model:} The HOF model is used to model the failure of a UE to hand over from its serving cell $c_0$ to its target cell $c^{\prime}$. The UE initiates a handover by using the contention-free random access (CFRA) resources to access the selected beam $b^{\prime}$ of target cell $c^{\prime}$. For successful random-access, it is a prerequisite that the SINR $\gamma_{c^{\prime},b^{\prime}}(m)$ of the target cell remains above the threshold $\gamma_\mathrm{out}$ during the RACH procedure. A HOF timer $T_\mathrm{HOF} $ = \SI{200}{ms} is started when the UE initiates the random-access towards the target cell $c^{\prime}$ and sends the RACH preamble. The RACH procedure is repeated until either a successful RACH attempt is achieved or $T_\mathrm{HOF}$ expires. A UE only succeeds in accessing the target cell if the SINR  $\gamma_{c^{\prime},b^{\prime}}(m)$ remains above the threshold $\gamma_\mathrm{out}$ and as such a successful HO is declared. A HOF is declared if the timer $T_\mathrm{HOF}$ expires and the UE fails to access the target cell, i.e., $\gamma_{c^{\prime},b^{\prime}}(m)<\gamma_\mathrm{out}$ for the entire duration that the HOF timer runs. The UE then performs connection re-establishment to a new cell (possibly the previous serving cell) and this procedure contributes to additional signaling overhead and signaling latency \cite{b1}. 

\textit{RLF Model:} The RLF model is used to model the failure of a UE while it is in its serving cell $c_0$. The UE keeps track of the radio link monitoring (RLM) SINR metric $\bar{\gamma}_\mathrm{RLM}$, which is an average of the downlink SINR measurements of the serving cell $\gamma_{c_0,b_0}$. An RLF timer $T_\mathrm{RLF}$ = \SI{1000}{ms} is started when the RLM SINR $\bar{\gamma}_\mathrm{RLM}$ of the serving cell $c_0$ drops below $\gamma_\mathrm{out}$, and if the timer $T_\mathrm{RLF}$ expires an RLF is declared. The UE then initiates connection re-establishment. While the timer $T_\mathrm{RLF}$ runs, the UE may recover before declaring an RLF if the SINR $\bar{\gamma}_\mathrm{RLM}$ exceeds a second SINR threshold defined as $\gamma_\mathrm{in}$ = \SI{-6}{dB}, where $\gamma_\mathrm{in} > \gamma_\mathrm{out}$ \cite{b1}. If the beam failure recovery \cite[pp. 2-3]{b17} process fails the UE also declares an RLF and this is also taken into account in the RLF model.

\subsection{Hand Blockage Cases}
\label{Subsec3.2}

\begin{figure*}[!t]    \begin{subfigure}{0.2\textwidth}        \centering        \includegraphics[width=1\textwidth]{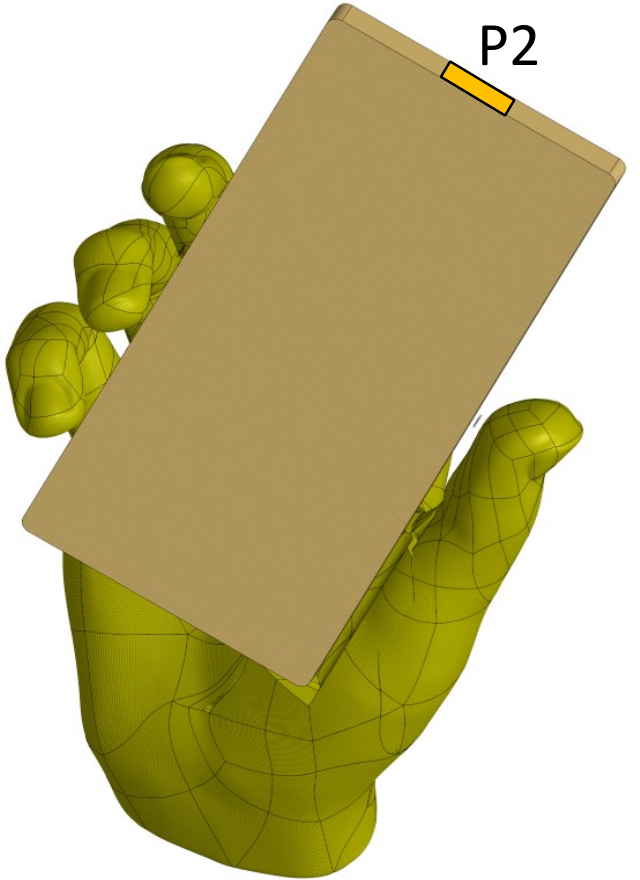}               \caption{Front view of the MPUE.}        \label{fig:Fig4a}    \end{subfigure}\hfill    
\begin{subfigure}{0.2\textwidth}        
\centering        \includegraphics[width=1\textwidth]{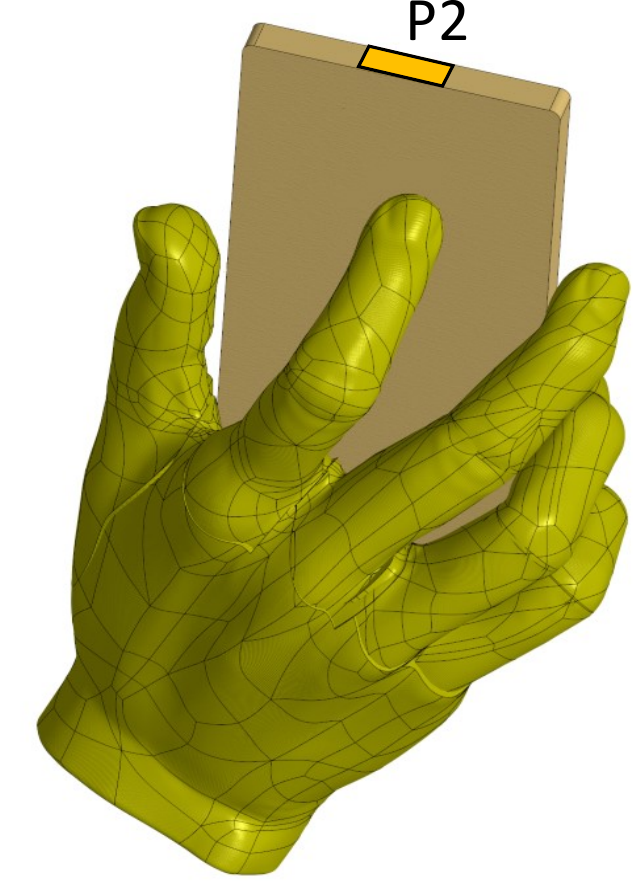}              \caption{Rear view of the MPUE.}        \label{fig:Fig4b}    \end{subfigure}\hfill 
\begin{subfigure}{0.2\textwidth}        
\centering        \includegraphics[width=1\textwidth]{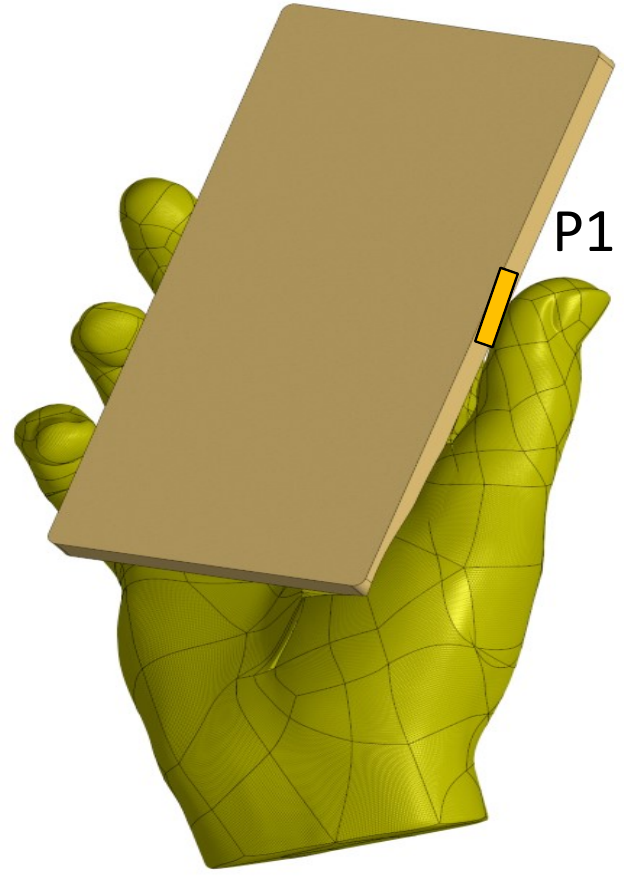}                \caption{P1 completely blocked by the thumb.}        \label{fig:Fig4c}    \end{subfigure}\hfill 
\begin{subfigure}{0.2\textwidth}        
\centering        \includegraphics[width=1\textwidth]{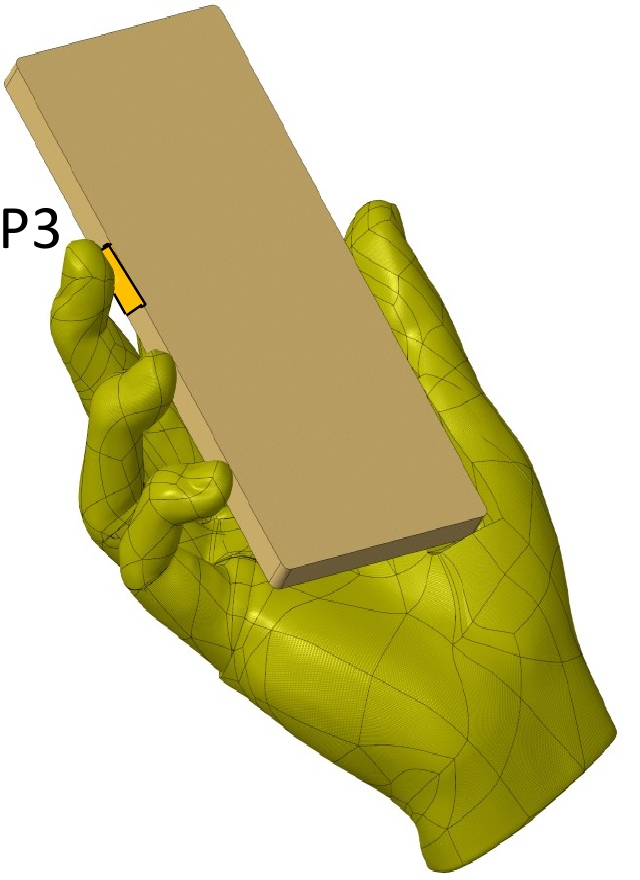}                \caption{P3 completely blocked by the middle finger.}        \label{fig:Fig4d}    \end{subfigure}\hfill 
\caption{The hand phantom model defined in \textit{\textbf{CST Studio Suite}} \cite{b16} for the MPUE in edge design is based on the CTIA wide-grip \cite[pp. 364]{b15}. It considers a \SI{1}{mm} air gap between the UE body and the finger (or thumb).}    \label{fig:Fig4}    \end{figure*}

A right-handed grip is considered for an MPUE that is in \textit{edge} design with three directional panels. As discussed in \Cref{Subsec2.1}, UE dimensions are taken as 15.7~cm $\times$ 7.6~cm $\times$ \SI{1.0}{cm}, which corresponds to a CTIA wide-grip for both talk and data mode. This corresponds to a loose hand grip with an air gap of \SI{1}{mm} between the UE body and fingers. As shown in Fig.\,\ref{fig:Fig4a} and  Fig.\,\ref{fig:Fig4b}, the CTIA wide-grip implies that panel 2 (P2) always is completely unblocked by the hand. P1 is on the right edge of the UE and depending on the positioning of the thumb it may be completely blocked or unblocked by the thumb, as shown in Fig.\,\ref{fig:Fig4c}. This simplification is drawn from the fact that directional antenna panels are rapidly being miniaturized as the technology matures \cite{b31}. The same can be said for P3 which is on the left edge of the UE and it may be completely blocked or unblocked by the middle finger, as shown in Fig.\,\ref{fig:Fig4d}.

The antenna element radiation patterns are generated in \textit{CST Studio Suite} for each of the three directional MPUE panels. Six different hand blockage cases can be defined, which are summarized in \Cref{Table2}. The free-space case (Case I) considers no hand placement around the UE and takes into account form factor considerations and reflection associated gains from the UE surface \cite{b10}. For P1 and P3, when the UE is held in hand either the panels experience no blockage (Case II) or both panels are completely blocked by the thumb and middle finger (Case III), or either only P1 is blocked by the thumb (Case IV) or P3 is blocked by the middle finger (Case V).  For each blockage case, a separate simulation study is considered where all UEs in the network are assumed to experience the type of blockage as defined by the blockage case. For comparison, the ideal Rx antenna element radiation pattern defined by 3GPP \cite{b30, b17} is considered (Case VI), where there is no hand blockage for any of the three panels. To the best of our knowledge, there has not been any study documenting the hand blockage probability for each of the respective panels in a real-world mobile environment for the CTIA wide-grip hand phantom model. Such a study could assign weights to the hand blockage probability of each panel, with the resultant performance analysis then being much closer to real-world scenarios. Taking this into account, it can be safely said the real-world mobility performance would lie in between the best (Case II) and worst hand blockage cases (Case III). Thereby, this work provides not only bounds for the expected real-world performance, but also the framework for evaluating the actual performance once statistical data about the hand blockage cases is available.

\begin{table}[!t]
\begin{center}
\caption{Hand blockage cases}
\label{Table2}
\begin{tabular}{| c | c | c | c |}
\hline
\bfseries Case & \bfseries P2  & \bfseries P1 & \bfseries P3 \\
\hline \hline
Case I: Free space   & Free- & Free- & Free- \\
with UE form factor & space & space & space \\
\hline
Case II: Fingers and thumb  & No &  No & No\\
not blocking any panel  & blockage & blockage & blockage\\
\hline
Case III: Fingers and thumb  & No &  Complete & Complete\\
blocking P1 and P3 & blockage & blockage & blockage\\

\hline
Case IV: Thumb   & No &  Complete & No \\
blocking P1 & blockage & blockage & blockage\\
\hline
Case V: Middle finger   & No &  No & Complete \\
blocking P3 & blockage & blockage & blockage\\
\hline
Case VI: 3GPP antenna radiation  & - & - & - \\
pattern with no blockage \cite{b30}  &  &  & \\
\hline 
\end{tabular}
\end{center}
 \vspace{-12pt}
\end{table}

\section{Hand Blockage Mobility Performance Evaluation} 
\label{Sec4}

In this section, the mobility performance of FCHO is compared with CHO in terms of KPIs for the different hand blockage  cases. The KPIs used for performance evaluation are explained below.

\subsection{KPI\lowercase{s}} \label{Subsec4.1}

\begin{itemize}

 \item
\textit{Successful Handovers:} Indicates the total number of successful HOs from the serving cell $c_0$ to the target cell $c^{\prime}$ in the network.

\item
\textit{Fast Handovers}: Indicates the sum of ping-pongs and short-stays in the network. A ping-pong is a successful handover followed by a handover back to the original cell within a very short time $T_\textrm{FH}$ \cite{b21}, e.g., 1~s. It is assumed that both these handovers could have been potentially avoided. A short-stay is a handover from one cell to another and then to a third one within $T_\textrm{FH}$. Here it is assumed that a direct handover from the first cell to the third one would have served the purpose. Although fast handovers are part of successful handovers, they are accounted for as a detrimental mobility KPI which adds unnecessary signalling overhead to the network.    

\item
\textit{Mobility Failures:} Indicates the sum of HOFs and RLFs in the network. These are described by the HOF and RLF models discussed earlier in \Cref{Subsec3.1}.

\end{itemize}
Successful handovers, fast handovers, and mobility failures are normalized to the total number of UEs $N_\mathrm{UE}$ in the network per minute and expressed as UE/min.

\begin{itemize}
\item
\textit{Outage:} Outage is defined as a time period when a UE is unable to receive data from the network due to a number of reasons. When the average downlink SINR of the serving cell $\gamma_{c_0, b_0}$ falls below $\gamma_\mathrm{out}$ it is assumed that the UE is unable to communicate with the network and, thus, in outage. Besides, if the HOF timer $T_{\mathrm{HOF}}$ expires due to a HOF or the RLF timer $T_{\mathrm{RLF}}$ expires due to an RLF, the UE initiates connection re-establishment and this is also accounted for as outage. A successful handover, although a necessary mobility procedure, also contributes to the total outage since the UE cannot receive any data during the time duration it is performing random access to the target cell $c^{\prime}$. This outage is modeled as relatively smaller (55~ms) than the outage due to connection re-establishment (180~ms) \cite{b21}. The outage KPI is denoted in terms of a percentage as

\begin{equation}
\label{Eq5} 
\textrm{Outage} \ (\%) = \frac{\sum_{\forall u}{\textrm{Outage duration of UE}} \ u} {N_\mathrm{UE} \ \cdot \ t_\mathrm{sim}} \ \cdot \ 100. 
\end{equation}
\end{itemize}

\subsection{Simulation Results} \label{Subsec4.2}
\begin{figure}[!t]    
\begin{subfigure}{0.5\textwidth}        
\centering        
\includegraphics[width=1\textwidth]{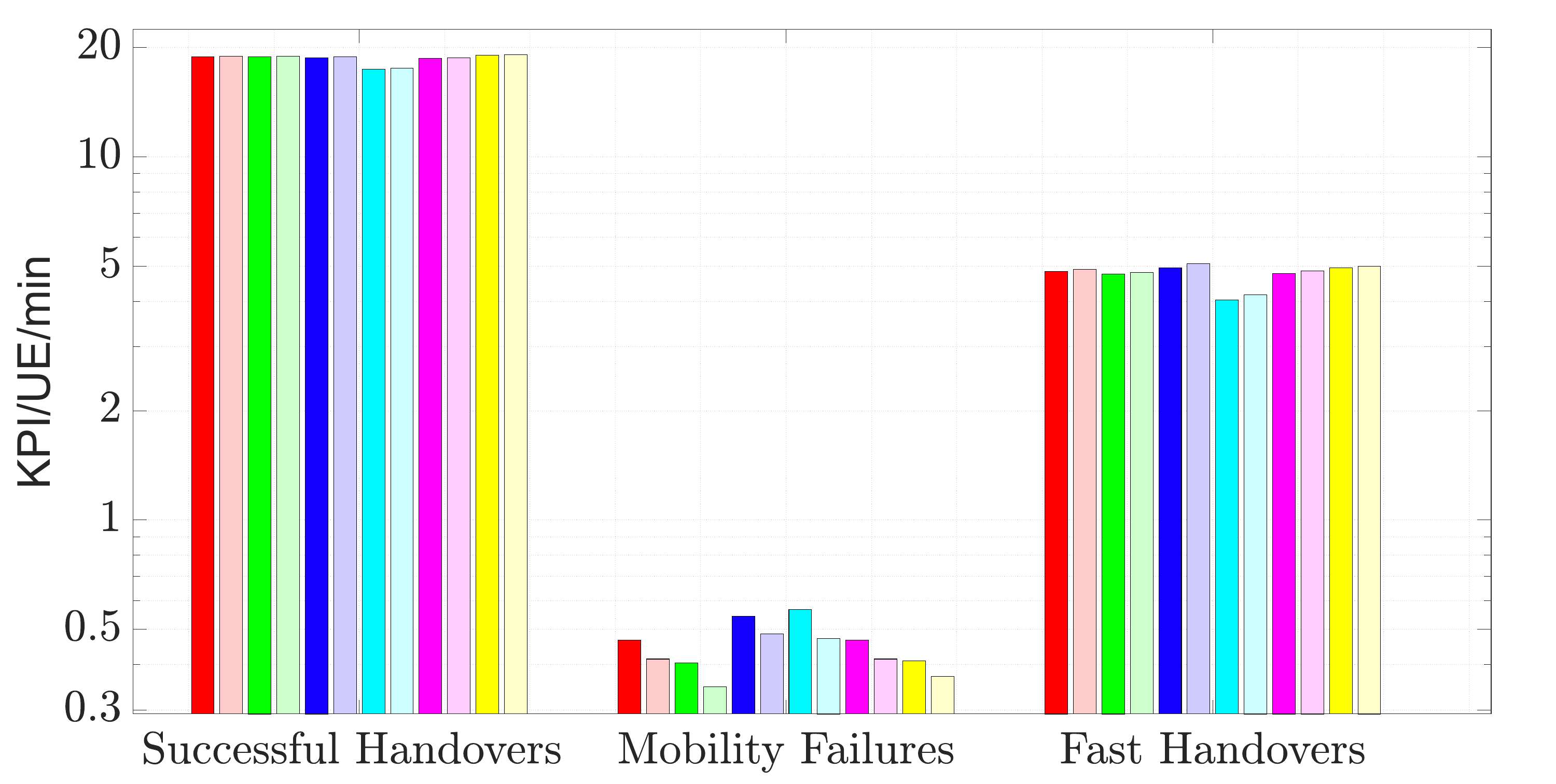}        \label{fig:Fig5a}
\vspace{-14pt}
\caption{Mobility KPIs.}        
\label{fig:Fig5a}    
\end{subfigure}\hfill    

\begin{subfigure}{0.5\textwidth}        
\centering        
\includegraphics[width=1\textwidth]{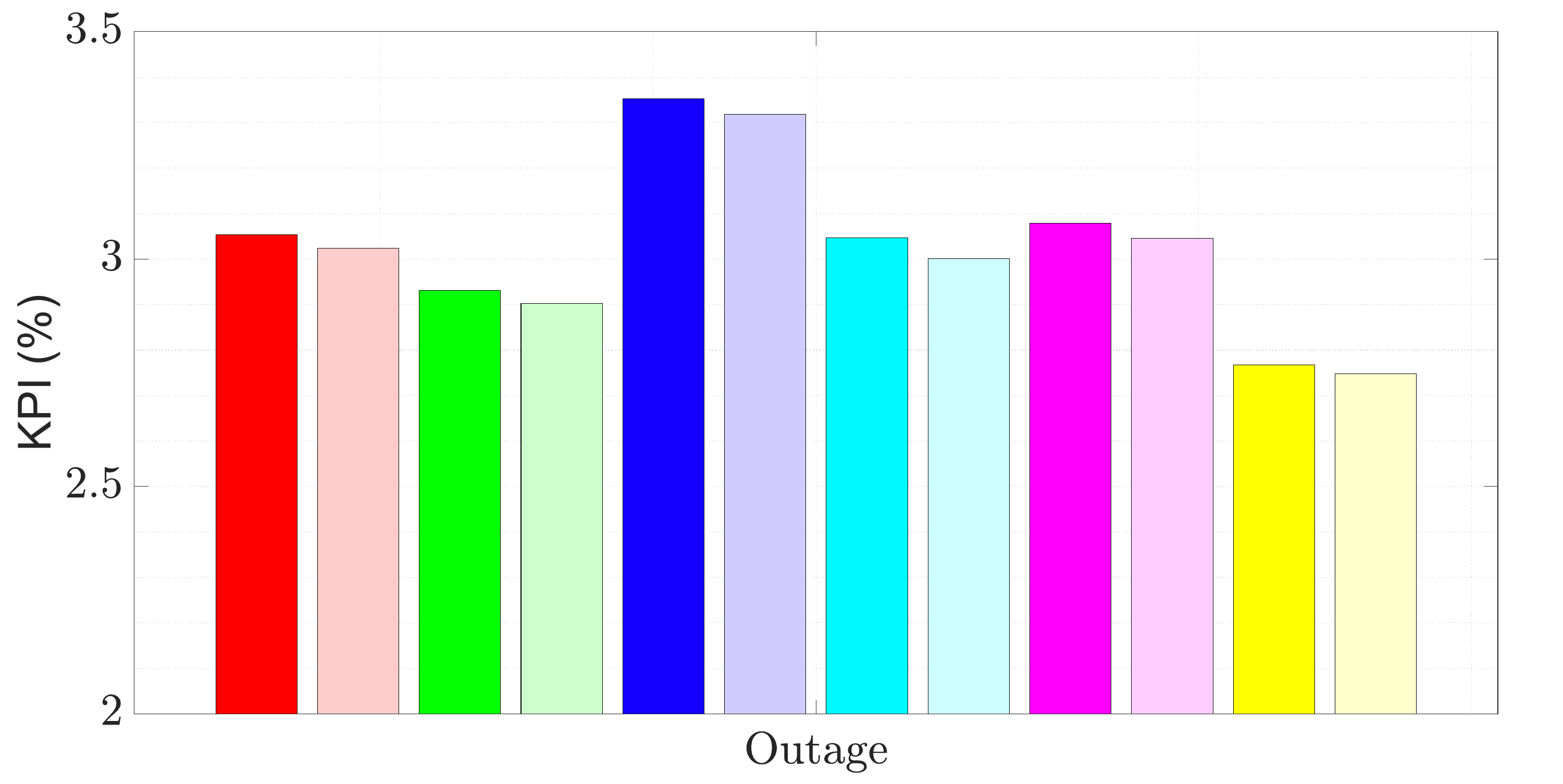}   
\label{fig:Fig5b}
\vspace{-14pt}
\caption{Outage KPI.} 
\label{fig:Fig5b}

\end{subfigure}\hfill

\begin{subfigure}{0.5\textwidth}     
\centering        
\includegraphics[width=1\textwidth]{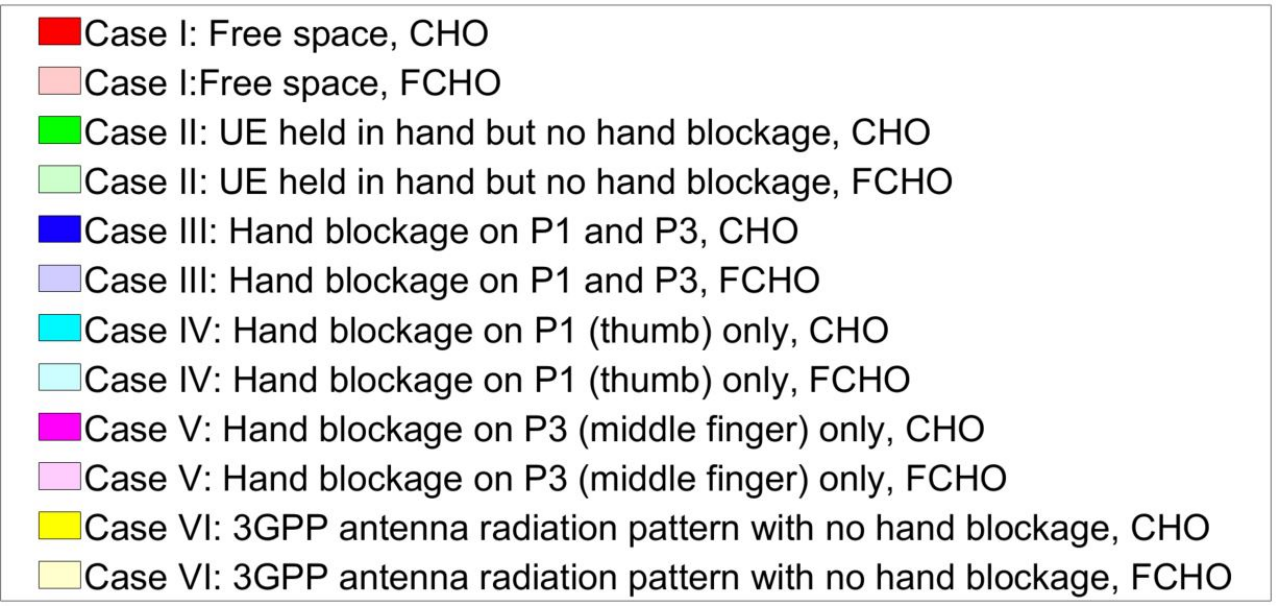}       
\vspace{-8pt}      
\end{subfigure}\hfill  
\vspace{-\baselineskip}
\caption{The mobility performance of FCHO compared with CHO for different hand blockage cases for the urban mobility scenario (\SI{60}{km/h}) for (a) mobility KPIs and (b) outage KPI.}
\label{fig:Fig5}   
\end{figure}

Fig.\,\ref{fig:Fig5} shows a comparison  of the mobility performance of FCHO with CHO for the urban mobility scenario (\SI{60}{km/h}) for each of the six hand blockage cases discussed in \Cref{Subsec2.2}.

The first key observation from Fig.\,\ref{fig:Fig5a} when FCHO is compared with CHO is that there is a decrease in mobility failures for each of the six blockage cases. For reference purposes, Case VI can be compared to the study in \cite{b6} wherein the hand blockage effect is not considered. For the free space case (Case~I, shown in red) the mobility failures decrease relatively by 11.2\% for FCHO when compared with CHO. For the panel blockage case on P1 and P3 (Case~III, shown in blue), the mobility failures decrease by 10.5\%. The other key observation is that hand-reflection associated gains can help to improve the mobility performance by decreasing mobility failures when none of the three MPUE panels are blocked. This can be seen if the free space case (Case~I, shown in red) is compared with the UE held in hand but no hand blockage (Case~II, shown in green) for their respective FCHO cases, where the mobility failures reduce substantially by 16.0\%. This was also one of the main conclusions of \cite{b11} where a loose hand grip with a similar air gap between the UE body and fingers was considered. If we compare the free-space case (Case~I, shown in red) with the 3GPP antenna radiation pattern with no hand blockage (Case~VI, shown in yellow) it is also seen that the form factor considerations and reflections from UE surfaces in a real-world mobile environment would contribute to more mobility failures because the antenna element directional gain for each of the MPUE panels is reduced in certain directions. 

\begin{figure*}[!t]    \begin{subfigure}{0.45\textwidth}        \centering        \includegraphics[width=1\textwidth]{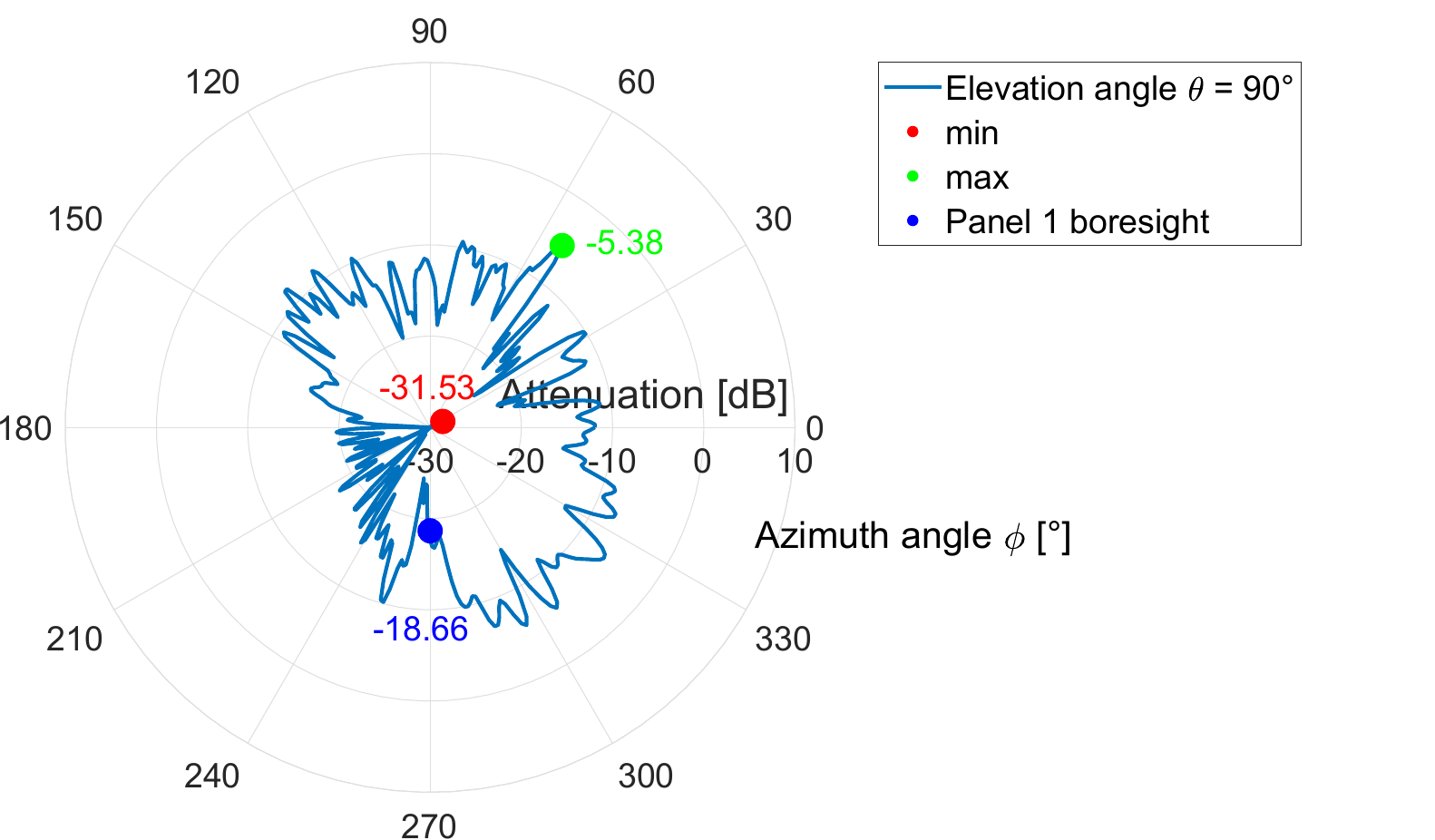}              \caption{Complete blockage on panel 1 (P1).}        \label{fig:Fig7a}    \end{subfigure}\hfill    
\begin{subfigure}{0.45\textwidth}        
\centering        \includegraphics[width=1\textwidth]{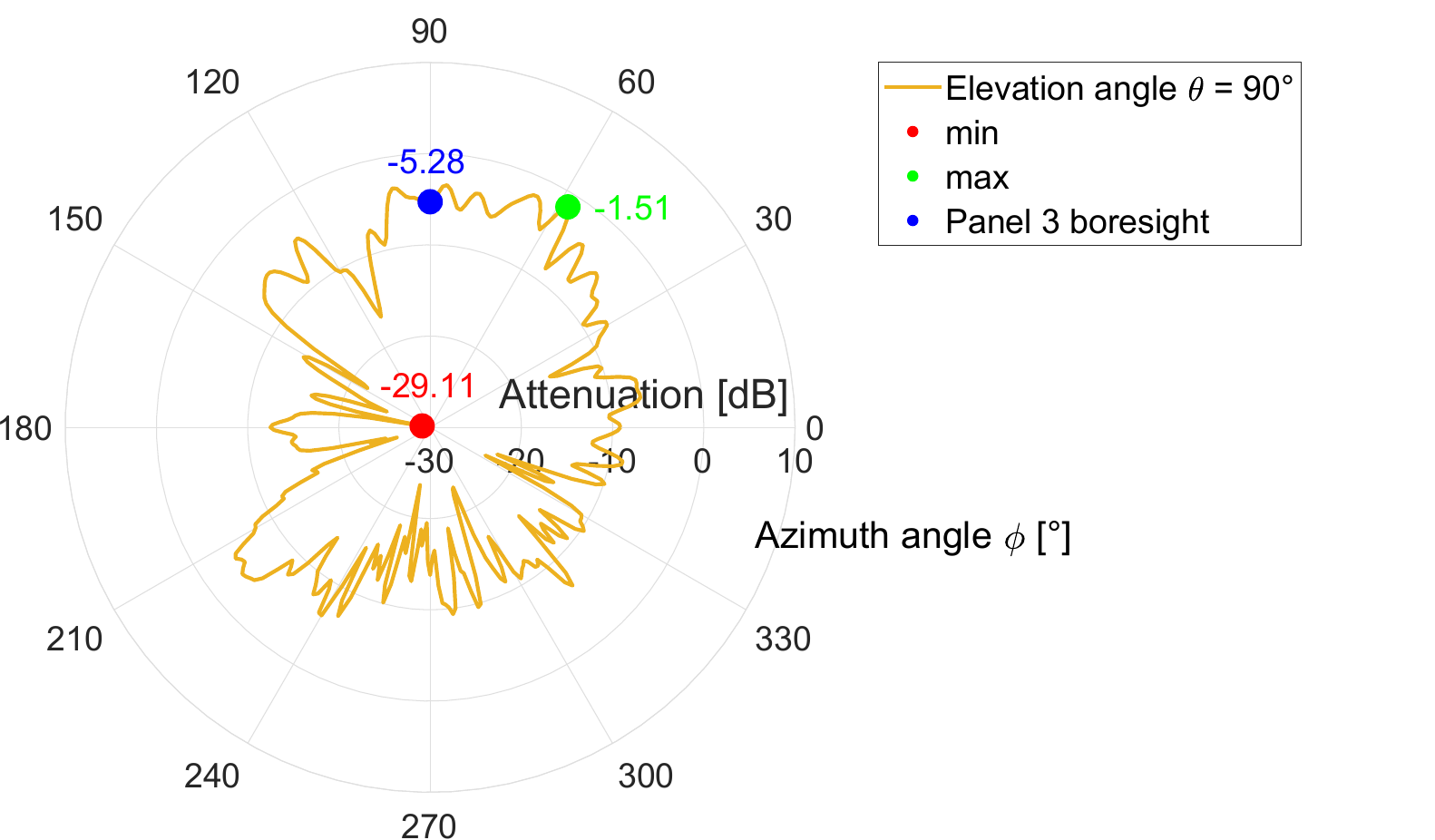}               \caption{Complete blockage on panel 3 (P3).}        \label{fig:Fig7b}    \end{subfigure}\hfill 
\caption{Antenna element radiation pattern (taken for elevation cut at $\theta=90^{\circ}$) (a) complete blockage on panel 1 (P1) and (b) complete blockage on panel 3 (P3).
}   \label{fig:Fig7}    \vspace{-14pt}\end{figure*}

It can also be visualized that there is an asymmetry between the blockage induced by the thumb on P1 (Case IV, shown in cyan) and the middle finger on P3 (Case~V, shown in magenta). If Case~IV is compared with Case~V for FCHO, it is seen that mobility failures in the latter are 12.1\% lower. Correspondingly the successful handovers are 6.7\% higher. This stems from the fact that the blockage effect of the thumb (Case~IV) is greater than that of the middle finger (Case~V). This is illustrated in Fig.\,\ref{fig:Fig7}, where it is seen that the attenuation at the boresight of P1 ($\phi = 270$) blocked by the thumb  in Fig.\,\ref{fig:Fig7a} is \SI{-18.66}{dB} whereas the corresponding attenuation at the boresight of P3 ($\phi = 90$) blocked by the middle finger in Fig.\,\ref{fig:Fig7b} is \SI{-5.28}{dB}. The effect of hand reflections is also visible since the maximum gain is seen not at the respective boresights but along other directions for both Fig.\,\ref{fig:Fig7a} and Fig.\,\ref{fig:Fig7b}. \textit{CST Studio Suite} considers the exact positioning of the hand and since the thumb runs parallel to P1 in Fig.\,\ref{fig:Fig4c} as compared to the middle finger which is at an angle to P3 in Fig.\,\ref{fig:Fig4d}, the blockage effect on P1 is greater. When Case~III  is compared with Case~V, it is seen that fast handovers (and consequently successful handovers) are greater in Case~III. This is because when both panels are blocked in Case~III, the UE may be forced to switch to P2 as its serving panel \cite[pp. 3]{b6} and the unreliable L3 RSRP measurements lead to more of such fast handovers.

Lastly, it can be observed in Fig.\,\ref{fig:Fig5b} that when FCHO is compared with CHO the outage decreases for all blockage cases on account of lower mobility failures. One of the main conclusions of \cite{b6} was that some mobility failures in CHO may translate into fast handovers in FCHO because such mobility failures are addressed in cell border regions, where the probability of fast handovers is very high. This can be observed in Fig.\,\ref{fig:Fig5a} where for FCHO the fast handovers increase for all of the six hand blockage cases that are considered in this study. It can also be observed that the increase in outage due to an increase in fast handovers with FCHO as compared to CHO (2.7\% for Case~III where both P3 and P1 are blocked, shown in blue) is negated by the  decrease in mobility failures (10.5\% for Case~III).  In \Cref{Subsec4.1} it was already discussed that the outage contribution of mobility failures is almost 4 times that of fast handovers. It can also be observed in Fig. \,\ref{fig:Fig5b} that the worst outage is experienced for Case~III (3.3\% for FCHO) since it experiences more mobility failures and successful handovers in total than the other blockage cases.

\begin{figure}[!t]    
\begin{subfigure}{0.5\textwidth}        
\centering        
\includegraphics[width=1\textwidth]{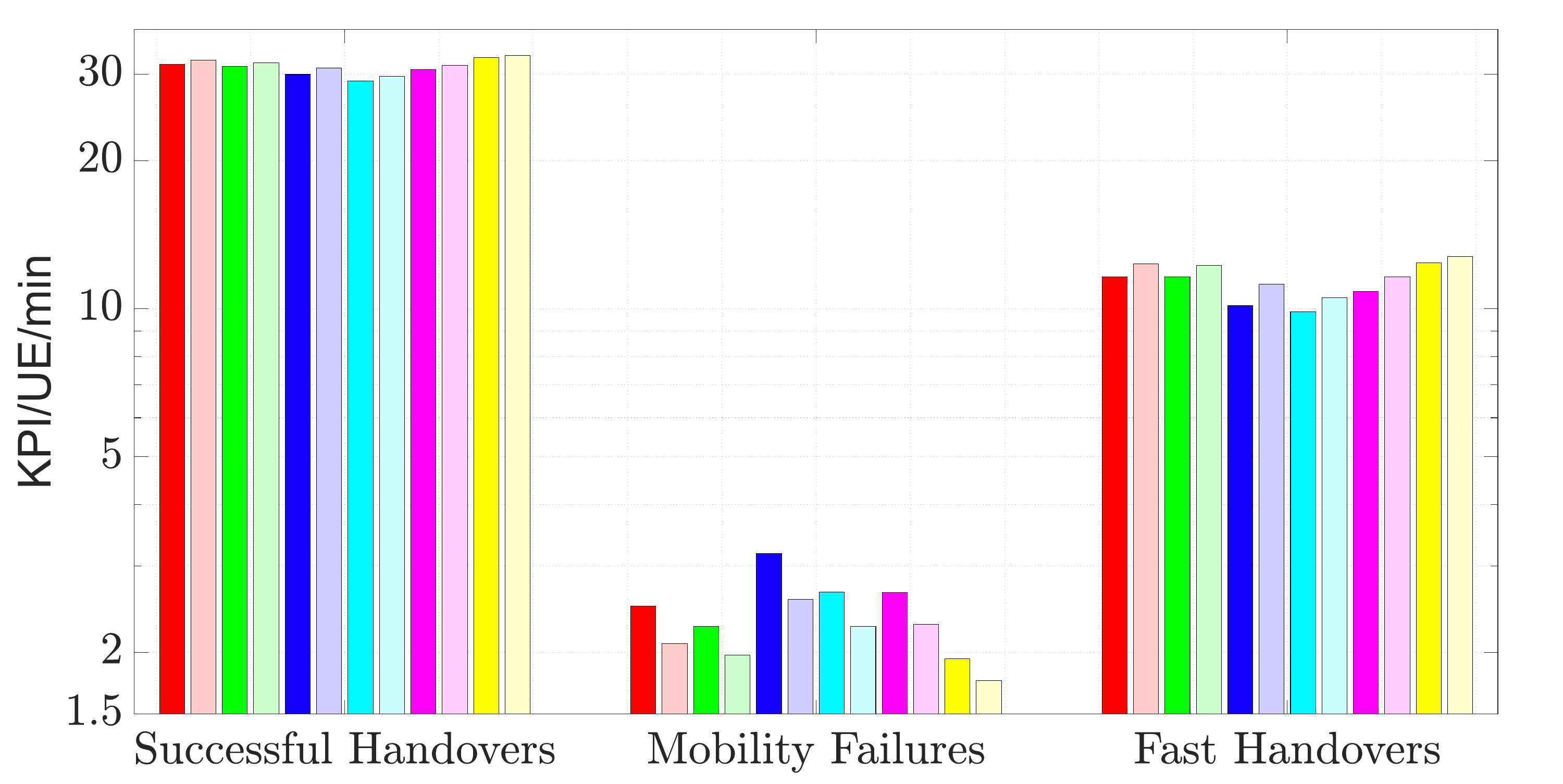}        \label{fig:Fig6a}
\vspace{-14pt}
\caption{Mobility KPIs.}        
\label{fig:Fig6a}    
\end{subfigure}\hfill    

\begin{subfigure}{0.5\textwidth}        
\centering        
\includegraphics[width=1\textwidth]{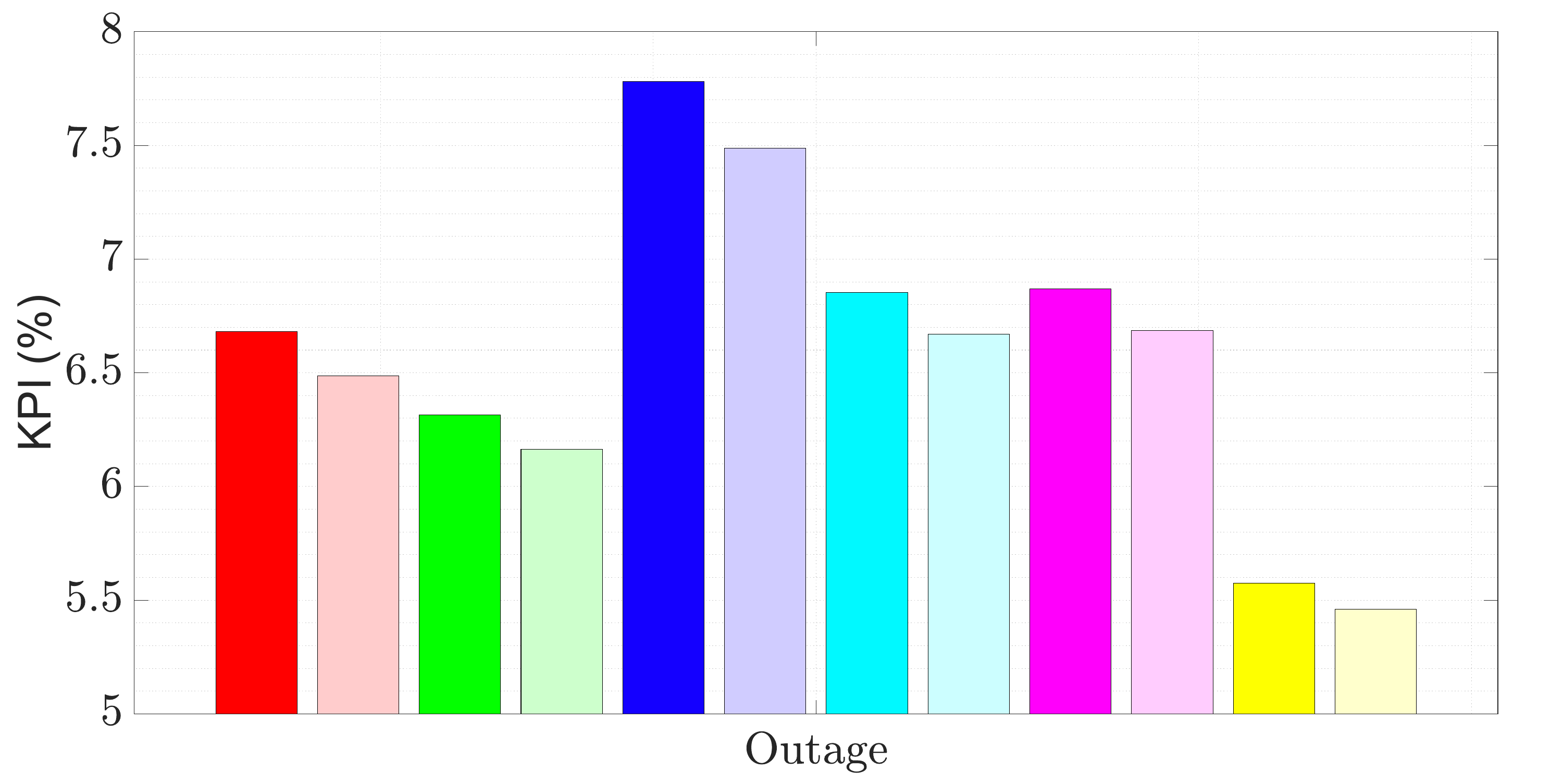}  
\label{fig:Fig6b}
\vspace{-14pt}
\caption{Outage KPI.}     
\label{fig:Fig6b}
\end{subfigure}\hfill

\begin{subfigure}{0.5\textwidth}     
\centering        
\includegraphics[width=1\textwidth]{Fig5c.pdf}        
\vspace{-8pt}  
\end{subfigure}\hfill  
\vspace{-1.4\baselineskip}
\caption{The mobility performance of FCHO compared with CHO for different hand blockage cases for the highway mobility scenario (\SI{120}{km/h}) for (a) mobility KPIs and (b) outage KPI.}
\label{fig:Fig6}   
\end{figure}

The mobility performance of FCHO is compared with CHO for the highway mobility scenario in Fig.\,\ref{fig:Fig6}. At \SI{120}{km/h} mobility is more challenging because of greater temporal variations in the signal RSRPs due to dominant fast fading. Additionally, the UEs traverse more cell boundaries compared to \SI{60}{km/h} and therefore the probability of mobility failures also increases. However, this also means that FCHO can be more beneficial at higher UE speeds in terms of addressing these mobility problems, which are now aggravated due to hand blockage. For the free space case (Case~I, shown in red) shown in  Fig.\,\ref{fig:Fig6a} the mobility failures decrease relatively by 16.1\% for FCHO when compared with CHO. For the panel blockage case on P1 and P3 (Case~III, shown in blue), the mobility failures decrease by 19.3\%. The asymmetry between blockage induced by the thumb on P1 (Case~IV, shown in cyan) and the middle finger on P3 (Case~V, shown in magenta) is not visible in terms of  mobility failures (which are nearly equal) but successful handovers in Case~V for FCHO are 5.58\% higher. Lastly, in Fig.\,\ref{fig:Fig6b} it can be observed that as seen in Fig.\,\ref{fig:Fig5b}, outage decreases for FCHO for all blockage cases and the worst outage is experienced for Case~III (7.5\% for FCHO).  

\begin{figure}[!t]    
\begin{subfigure}{0.5\textwidth}        
\centering        
\includegraphics[width=1\textwidth]{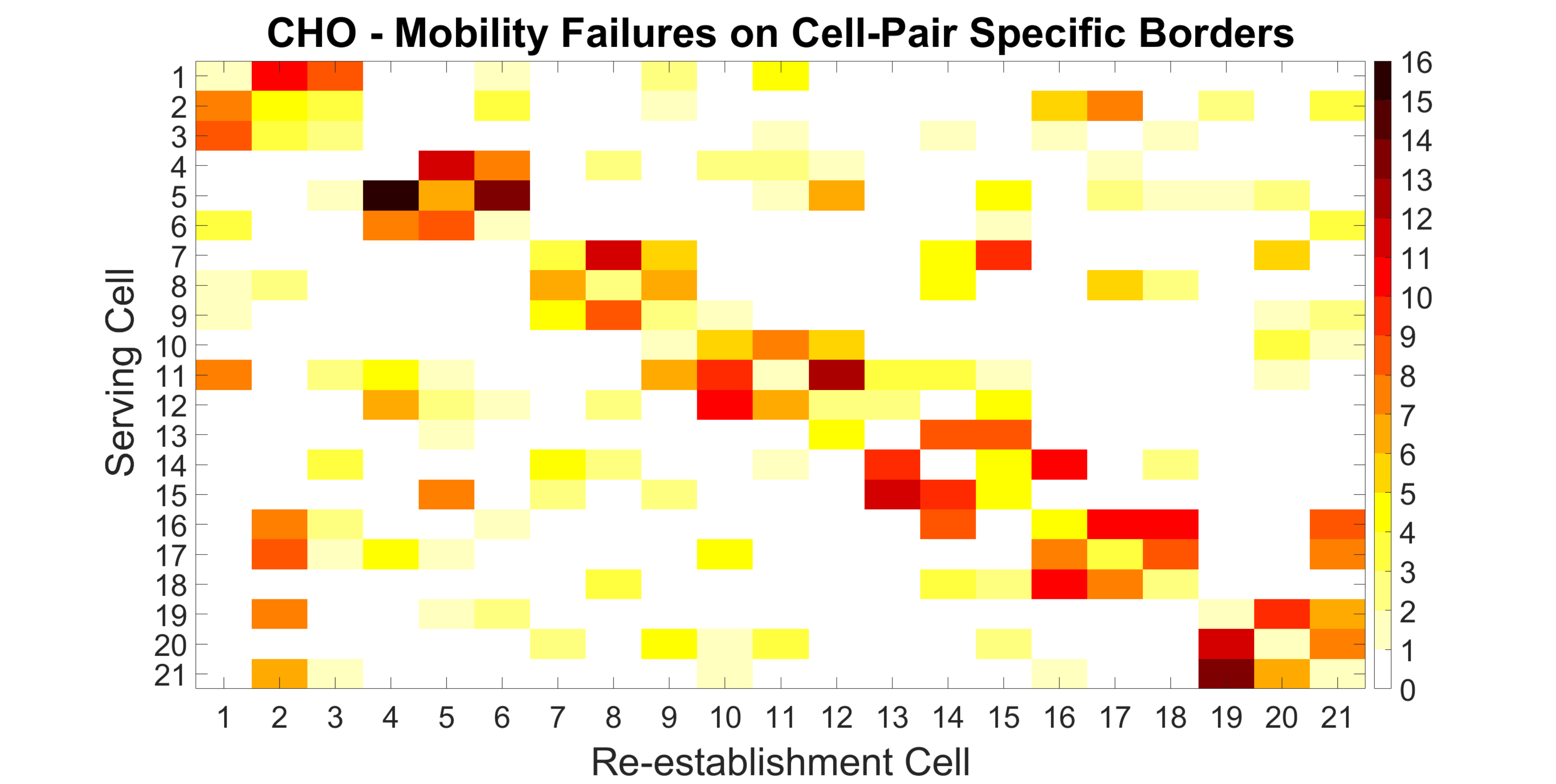}        \label{fig:Fig7.1a}
\vspace{-14pt}
\caption{Number of mobility failures on each cell-pair specific border.}        
\label{fig:Fig7.1a}    
\end{subfigure}\hfill    

\begin{subfigure}{0.5\textwidth}        
\centering        
\includegraphics[width=1\textwidth]{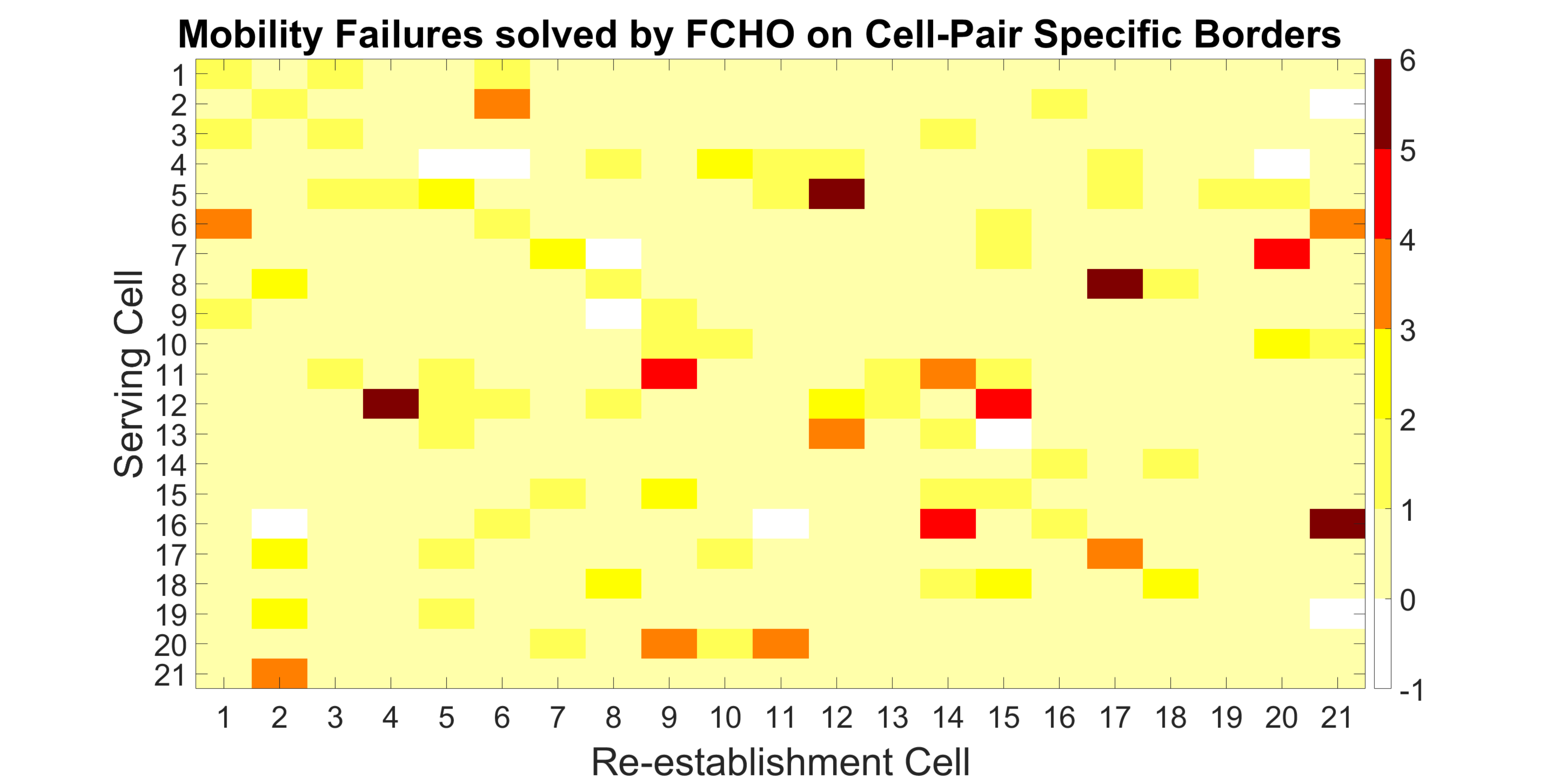}  
\label{fig:Fig7.1b}
\vspace{-14pt}
\caption{Number of mobility failures solved by FCHO on each cell-pair specific border.}     
\label{fig:Fig7.1b}
\end{subfigure}\hfill 

\begin{subfigure}{0.5\textwidth}        
\centering        
\includegraphics[width=1\textwidth]{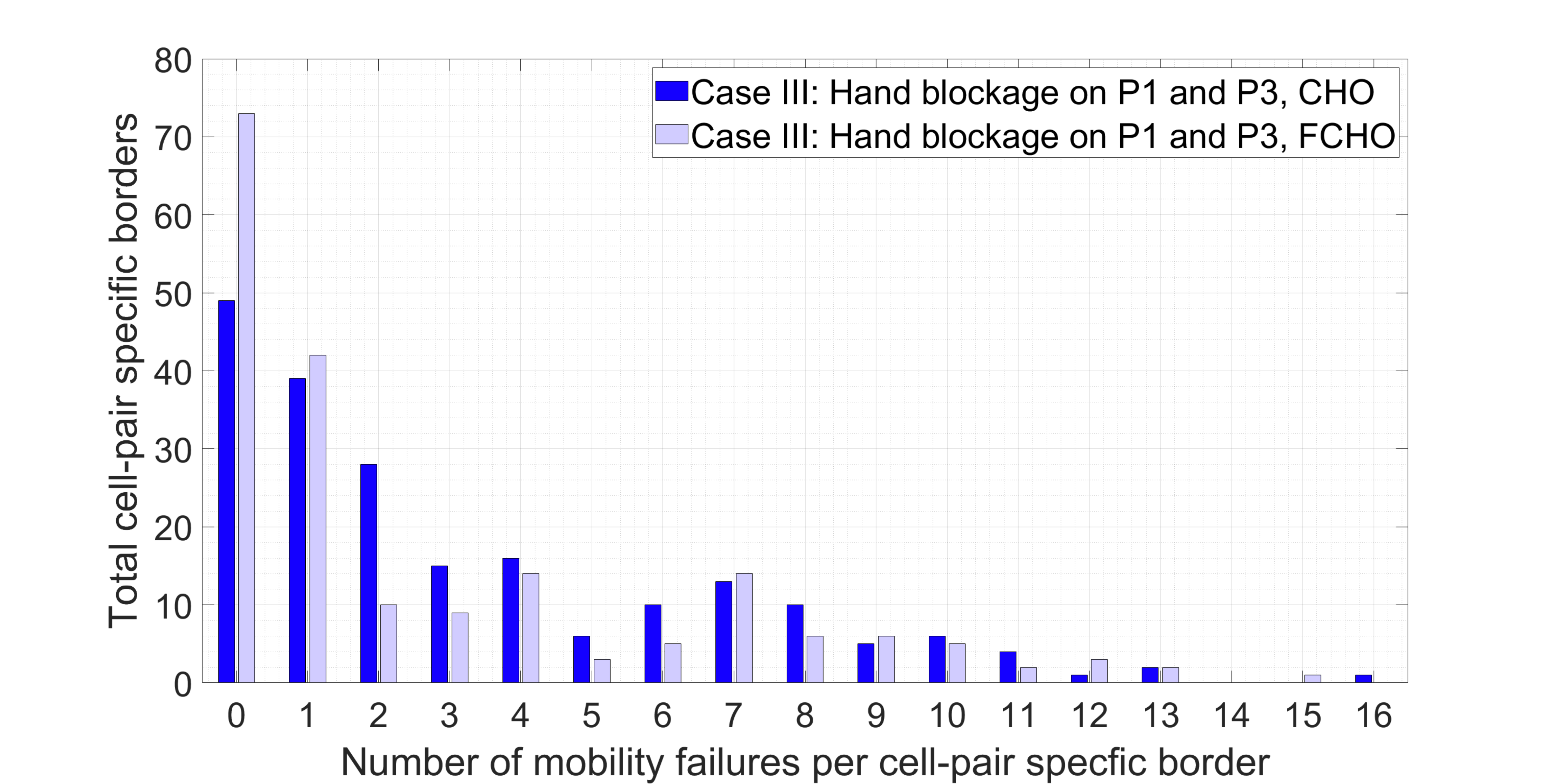}  
\label{fig:Fig7.1c}
\vspace{-14pt}
\caption{A histogram distribution of the number of mobility failures per cell-pair specific border.}     
\label{fig:Fig7.1c}
\end{subfigure}\hfill 
\caption{A comparison of the mobility failures between CHO and FCHO in terms of cell-pair specific borders for Case III of \Cref{Table2} and the highway mobility scenario (\SI{120}{km/h}).}
\label{fig:Fig7.1}   
\end{figure}

It is also useful to analyse the mobility failures for cell-pair specific borders shown in Fig.\,\ref{fig:Fig3}. This can be seen in Fig.\,\ref{fig:Fig7.1} where a comparison is made between CHO and FCHO for Case III of \Cref{Table2} for the highway mobility scenario (120 km/h). It can be visualized from the colormap shown in Fig.\,\ref{fig:Fig7.1a} that for a given serving cell, there are only certain cell-pair specific borders that are problematic on account of their geographical neighbor relations, e.g., the cell border between cell 5 and 4 which experiences sixteen such failures where the UE fails in serving cell 5 and re-establishes to cell 4. It can also be seen that most cell-pair specific borders have no mobility failures on account of having no geographical neighbor relation, which also includes coverage islands and the wrap-around effect mentioned in \Cref{Subsec3.1}. It can be seen in Fig.\,\ref{fig:Fig7.1b} that FCHO addresses mobility failures for most of these cell-pair specific borders, e.g., the border between cell 12 and 4 where the difference is six mobility failures (six mobility failures reduced to zero). It can also be visualized that in a few cell borders, FCHO could lead to more mobility failures because a handover attempt could be made to a retained cell for which the radio link is degrading due to UE movement or changing radio link conditions. However, the overall benefit of FCHO is clearly visible in Fig.\,\ref{fig:Fig7.1c}, where it is now seen that there are twenty-four cell-pair specific borders (forty-nine and seventy-three cell-pair specific borders with zero mobility failures, respectively, for CHO and FCHO) with a geographical neighbor relation where FCHO reduces the number of mobility failures to zero. 

\section{Resource Reservation Time Optimization in FCHO} 
\label{Sec5}

In this section, the resource reservation problem in FCHO is discussed in detail. Further on, three different FCHO optimization approaches that address this resource reservation problem are explained. 

\subsection{Resource Reservation Problem in FCHO} \label{Subsec5.1}

 One of the main conclusions of our earlier studies \cite[pp. 6-7]{b6} was that while FCHO is advantageous over CHO in terms of reducing both the mobility failures and signaling overhead, the resource reservation time in FCHO is significantly higher due to the retention of the conditional configurations of the prepared target cells after each handover. Resource reservation time can be defined as the time duration for which a prepared target cell $c^{\prime}$ reserves resources for a particular UE $u$. It can be formulated as
\begin{equation}
\label{Eq6} 
    t_{c^{\prime},u}^{R} (n^\mathrm{prep}_{c^{\prime}}) = \sum_m \chi_{n^\mathrm{prep}_{c^{\prime}}(m)}(u) \Delta t,
\end{equation}
where $n^\mathrm{prep}_{c^{\prime}}$ contains the lists of prepared UEs of cell $c^{\prime}$ over time as defined in (\ref{Eq4.6}), $\chi$ is the indicator function and  $\Delta t$ is the time step size defined in \Cref{Table1}.
 
 In FCHO if a UE maintains a preparation for a target cell it implies that the prepared target cell $c^{\prime}$ needs to reserve the resources for the particular UE in question till they are otherwise released by the network through either the CHO release condition in (\ref{Eq2}) or replaced by another cell through the CHO replace condition in (\ref{Eq3}). Unlike CHO, the resources for the prepared cells are not explicitly released after a handover. This leads to longer resource reservation time $t_{c^{\prime},u}^{R} (n^\mathrm{prep}_{c^{\prime}})$ in FCHO. Some examples of such resources are contention-free random access preambles, UE identifiers, radio resources for guaranteed bit rate radio bearers, and even hardware resources such as buffers to receive the packets forwarded early from the serving cell \cite{b315}. 

Fig.\,\ref{fig:Fig8} shows the cumulative distribution function (CDF) of the resource reservation time in a prepared target cell for Case III of \Cref{Table2} and the highway mobility scenario (\SI{120}{km/h}). It is seen that at the 50\textsuperscript{th} percentile, the resource reservation time  $t_{c^{\prime},u}^{R} (n^\mathrm{prep}_{c^{\prime}})$ for CHO (shown in red) is \SI{0.37}{s} as compared to \SI{0.47}{s} for FCHO (shown in blue). For the 95\textsuperscript{th} percentile the resource reservation time is \SI{1.81}{s} for CHO and \SI{2.55}{s} for FCHO, leading to a difference of \SI{0.74}{s}. This clearly indicates that in FCHO the prepared target cells reserve resources for the UEs for a much longer time duration as compared to CHO. Resultantly, the capacity of the network nodes to admit UEs during handover or connection setup decreases, which can be critical, especially in high-load situations since it could lead to an increase in mobility failures in case of failure to prepare a target cell for handover. Therefore, the resource reservation time in FCHO needs to be optimized so as to decrease it and bring it as close as possible to that of CHO. Some earlier studies \cite{b316, b317, b318} have focused on the call blocking probability of new calls and forced termination probabilities of ongoing calls during handover, both due to the unavailability of idle resources on the network side. In this article, we focus on an admission control scheme where idle resources are always available and where all handover requests sent by the serving cell $c_0$ to the target cell $c^{\prime}$ are approved by $c^{\prime}$. This then implies that in (\ref{Eq4.6}) the upper bound value is $n_c^{\mathrm{max}} = N_\mathrm{UE}$. By studying the resource reservation time, however, we do take into account the required system capacity where the prepared target cells are released either due to a handover or due to the CHO release or replace conditions mentioned in (\ref{Eq2}) and (\ref{Eq3}), respectively.

\begin{figure}[!t]
\textit{\centering
\includegraphics[width = 1\columnwidth]{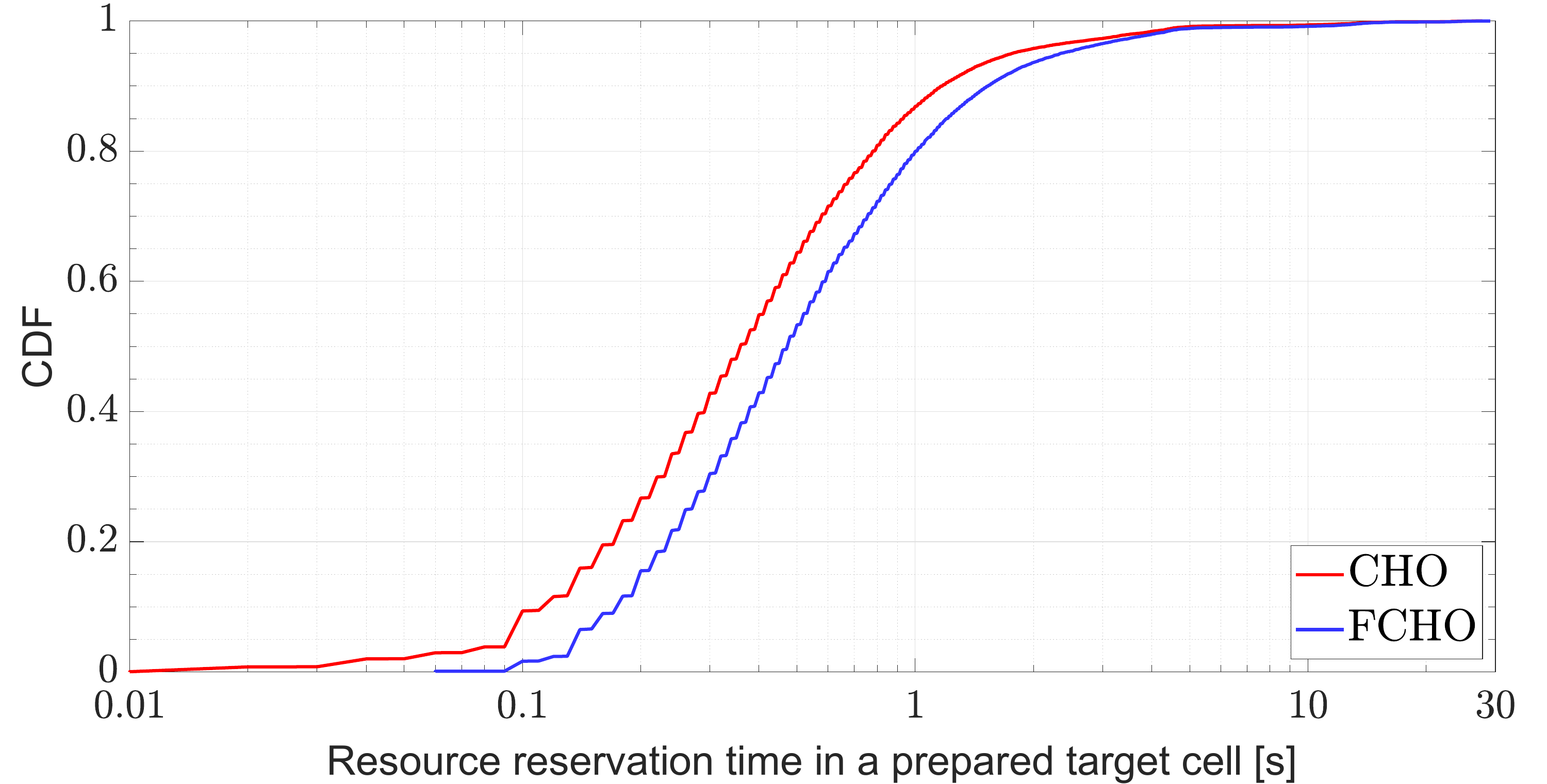}
\label{fig:Fig8}
\vspace{-\baselineskip}
\caption{A comparison of the resource reservation time in a prepared target cell between CHO and FCHO for Case III of \Cref{Table2} and the highway mobility scenario (\SI{120}{km/h}).} 
\label{fig:Fig8}} 
\end{figure}

\subsection{FCHO Resource Reservation Time Optimization Approaches} \label{Subsec5.2}

As outlined in \Cref{Subsec5.1}, we intend to improve the resource reservation time  $t_{c^{\prime},u}^{R} (n^\mathrm{prep}_{c^{\prime}})$ for FCHO. This improvement should not come at the cost of mobility failures or signaling overhead compared to CHO and FCHO, respectively. To that effect, we collect the lists of prepared UEs over time from all cells in the tuple $\mathbf{n}^{\mathrm{prep}}$ defined as
\begin{equation} \label{Eq7}
    \mathbf{n}^{\mathrm{prep}} = ( n_1^\mathrm{prep}, \dots, n_{N_\mathrm{cells}}^\mathrm{prep} ).
\end{equation}
As such, the optimization problem can be summarized as
%
\begin{subequations} \label{Eq8}
\begin{alignat}{2}
&\!\min_{\mathbf{n}^{\mathrm{prep}}}         &\qquad& t{^{R_{\mathrm{norm}}}} (\mathbf{n}^{\mathrm{prep}} )\\
&\text{subject to} &      & f(\mathbf{n}^{\mathrm{prep}}) \leq f_\mathrm{CHO},\\
&                  &      & s(\mathbf{n}^{\mathrm{prep}}) \leq  s_\mathrm{FCHO},
\end{alignat}
\end{subequations}
where $t{^{R_{\mathrm{norm}}}} (\mathbf{n}^\mathrm{prep})$ is the overall normalized resource reservation time in a prepared target cell $c^{\prime}$, $f_{\mathrm{CHO}}$ is the normalized number of mobility failures in CHO  (KPI/UE/min) and $s_{\mathrm{FCHO}}$ is the normalized signaling overhead in FCHO (KPI/UE/min). The normalized resource reservation time can be defined as 
\begin{equation} 
\label{Eq9} 
t{^{R_{\mathrm{norm}}}} (\mathbf{n}^\mathrm{prep}) = \frac{\sum\limits_{c^{\prime}}\sum\limits_{u} \sum\limits_m {t_{c^{\prime},u}^{R}} (n_{c^{\prime}}^{\mathrm{prep}} (m)) }{N_\mathrm{cells} \cdot N_\mathrm{UE} \cdot  t_\mathrm{sim}} .
\end{equation}

Three different algorithms following the paradigm of MRO are developed to optimize the resource reservation in FCHO. MRO has already been designated as one use case of self-organizing networks (SON) \cite{b32, b33, b34} in 3GPP \textit{Release 9} \cite{b20}. In line with that, the resource reservation optimization approach for the three algorithms considers handover statistics that are collected from the network and processed. Thereafter, based on those statistics the parameters of the CHO preparation event are adjusted in the network to yield the desired results. 

For the statistics collection and processing, Case III of \Cref{Table2} is simulated for FCHO with the pre-configured offsets outlined in \Cref{Table1} for the highway mobility scenario. This case has been chosen since it represents the most challenging mobility scenario of all the cases in \Cref{Table2} where both P1 and P3 experience hand blockage. Based on the network-wide statistics that are collected at the end of the simulation, a handover probability matrix $\boldsymbol{H}$ is defined. $\boldsymbol{H}$ is an $N_\mathrm{cells} \times N_\mathrm{cells}$ matrix, where $H_{ij}$ represents the observed handover probability from cell $i$ (serving cell) to cell $j$ (prepared target cell). Thus, each row of $\boldsymbol{H}$ represents the probability of handover from cell $i$ to each of the other cells in the network.

As for the parameter adjustment part, the aim is to make both the preparation of cells (defined through the CHO preparation condition in (\ref{Eq1})) and retention of prepared cell conditional configurations after each handover harder for cases where the observed handover probability from serving cell $c_0$ to prepared target cell $c^{\prime}$ is small. For this, three different optimization approaches are defined as follows:
\begin{enumerate}
\vspace{-0.2pt}
    \item \textit{FCHO optimization with block listing}: In this approach, the preparation of certain cells is blocked if they have an observed handover probability $H_{ij} \leq p_\mathrm{B}$, where $p_\mathrm{B}$ is the block listing probability and  $p_\mathrm{B}=0$. We term this as \textit{active} block listing since certain cells will never be prepared and are part of a block list. Additionally, immediately after a handover, the optimization algorithm releases those retained target cells that have an observed handover probability $H_{ij} \leq p_\mathrm{B}$ by discarding them and this is termed as \textit{reactive} block listing. The preparation of these cells was not previously blocked since the observed handover probability $H_{ij}$ from the previous serving cell $i$ to the retained cell $j$ is not equivalent to $p_\mathrm{B}$. In CHO, all of the previously prepared target cells are discarded whereas in FCHO none of the prepared target cells are explicitly discarded but are instead retained (in addition to the previous serving cell which also becomes a target cell \cite[pp.4]{b6}). This optimization mechanism leads to both the block listing as well as the release of unnecessary cell preparations that the UE will most likely not hand over to, thus saving both signaling overhead and resource reservation time without degrading the mobility performance in terms of failures.

    \item \textit{FCHO optimization with preparation offset reduction}: In this approach, the preparation of certain target cells, controlled by the CHO preparation condition in (\ref{Eq1}), is delayed by reducing the CHO preparation offset between cell $i$ (the serving cell) and cell $j$ (the prepared target cell), i.e., $o^\mathrm{prep}_{i, j}$. As such, $o^\mathrm{prep}_{i, j}$ for those cells that have $H_{ij} \leq p_\mathrm{R}$ is reduced from its present value of \SI{10}{dB} to \SI{7}{dB}, where $p_\mathrm{R}$ is the preparation offset reduction probability and $p_\mathrm{R}=0.12$. The value of $p_\mathrm{R}$ is taken as 0.12 because it is observed in the collected statistics on average 3/4 of the preparations for a given serving cell have $H_{ij}\leq0.12$.  Thus, it represents an optimal value in terms of optimizing mobility performance, signaling overhead, and resource reservation. The aim of delaying the preparation for cells with low observed handover probability $H_{ij}$ is to ensure that these cells are prepared only when preparation becomes absolutely necessary. As a result of this delay, the target cell preparations are either delayed or in some cases avoided because either the RSRP of the serving cell recovers or the RSRP of the target cell degrades and the CHO preparation condition in (\ref{Eq1}) is not fulfilled. In this optimization mechanism, both the signaling overhead and resource reservation time will be reduced but a minimal increase in mobility failures is to be expected since some essential target cell preparations which are needed for successful handover may be delayed.

    \item \textit{FCHO optimization with block listing and preparation offset reduction}: In this approach, the aforementioned optimization mechanisms are combined. $o^\mathrm{prep}_{i, j}$ is now reduced for cells with observed handover probability between $0< H_{ij} \leq p_\mathrm{R}$ (instead of $H_{ij} \leq p_\mathrm{R}$) since the preparation of target cells is blocked for $H_{ij} \leq p_\mathrm{B}$, where $p_\mathrm{B}=0$. The preparation offset reduction probability is again taken as $p_\mathrm{R}= 0.12$. 
\end{enumerate}

The memory space complexity of these three different optimization approaches can be given as $\mathcal{O}(N_\mathrm{cells}^2)$. This is because it involves the formulation of the handover probability matrix $\boldsymbol{H}$, which has the dimension $N_\mathrm{cells} \times N_\mathrm{cells}$ and hence $N_\mathrm{cells}^2$ total entries. The CHO preparation offset $o^\mathrm{prep}_{i, j}$ which is modified depending on the handover probability $H_{ij}$ is also stored in a matrix with dimension $N_\mathrm{cells} \times N_\mathrm{cells}$. The time complexity is also given as $\mathcal{O}(N_\mathrm{cells}^2)$. It is pertinent to mention here that an increase in $N_\mathrm{cells}$ would not necessarily mean that the time and memory space complexity grows as given by the expressions, since the geographical neighbor relation for a given cell will not necessarily include all the other cells in the network as seen earlier in \Cref{Subsec3.1}. Hence, the given expressions serve as upper bounds for the time and memory space complexity of these optimization approaches.

\section{FCHO Resource Reservation Optimization Performance Analysis} \label{Sec6}

\begin{figure}[!t]    
\begin{subfigure}{0.5\textwidth}        
\centering        
\includegraphics[width=1\textwidth]{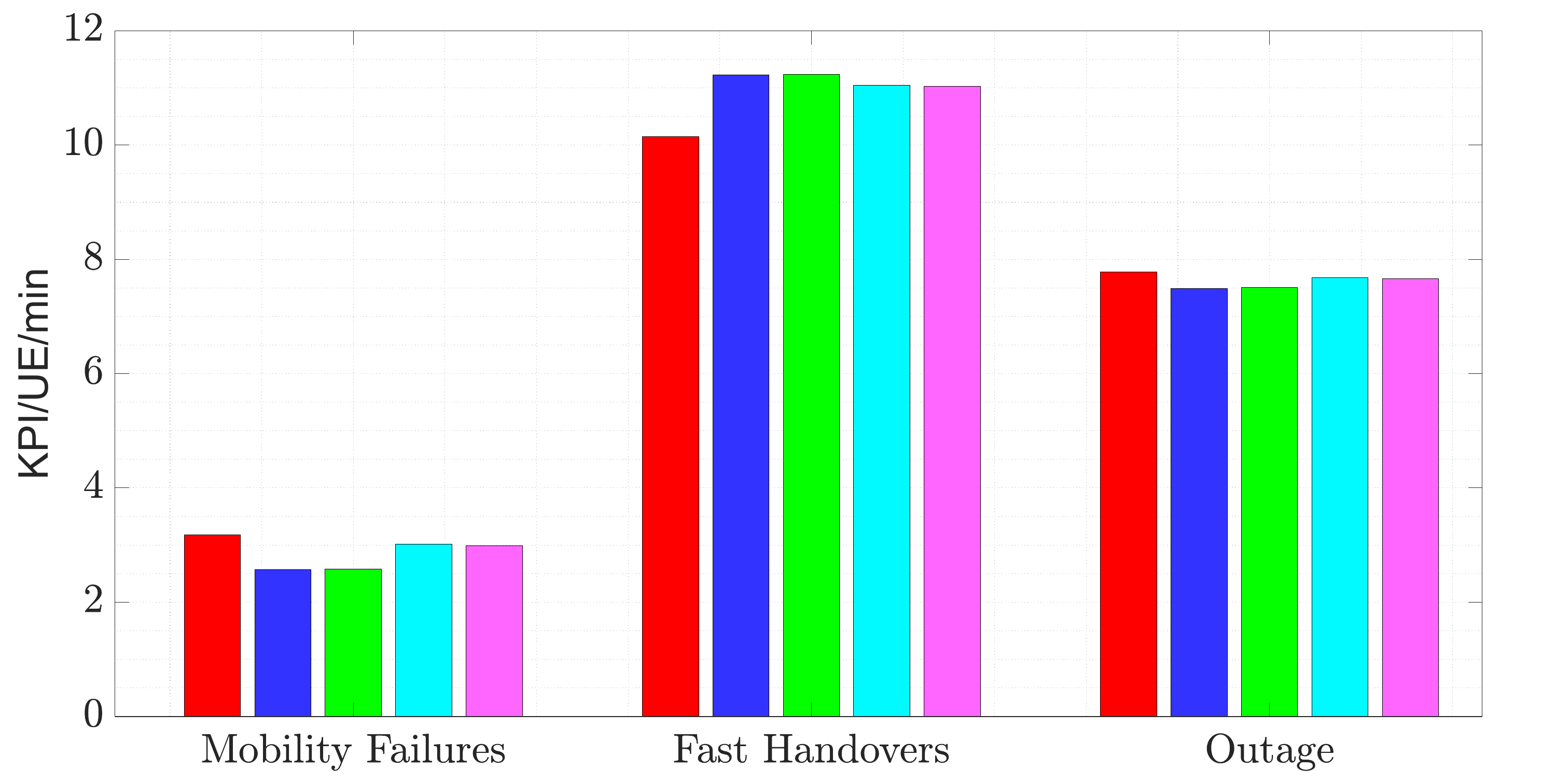}        \label{fig:Fig9a}
\vspace{-14pt}
\caption{Mobility and outage KPIs.}        
\label{fig:Fig9a}    
\end{subfigure}\hfill    

\begin{subfigure}{0.5\textwidth}        
\centering        
\includegraphics[width=1\textwidth]{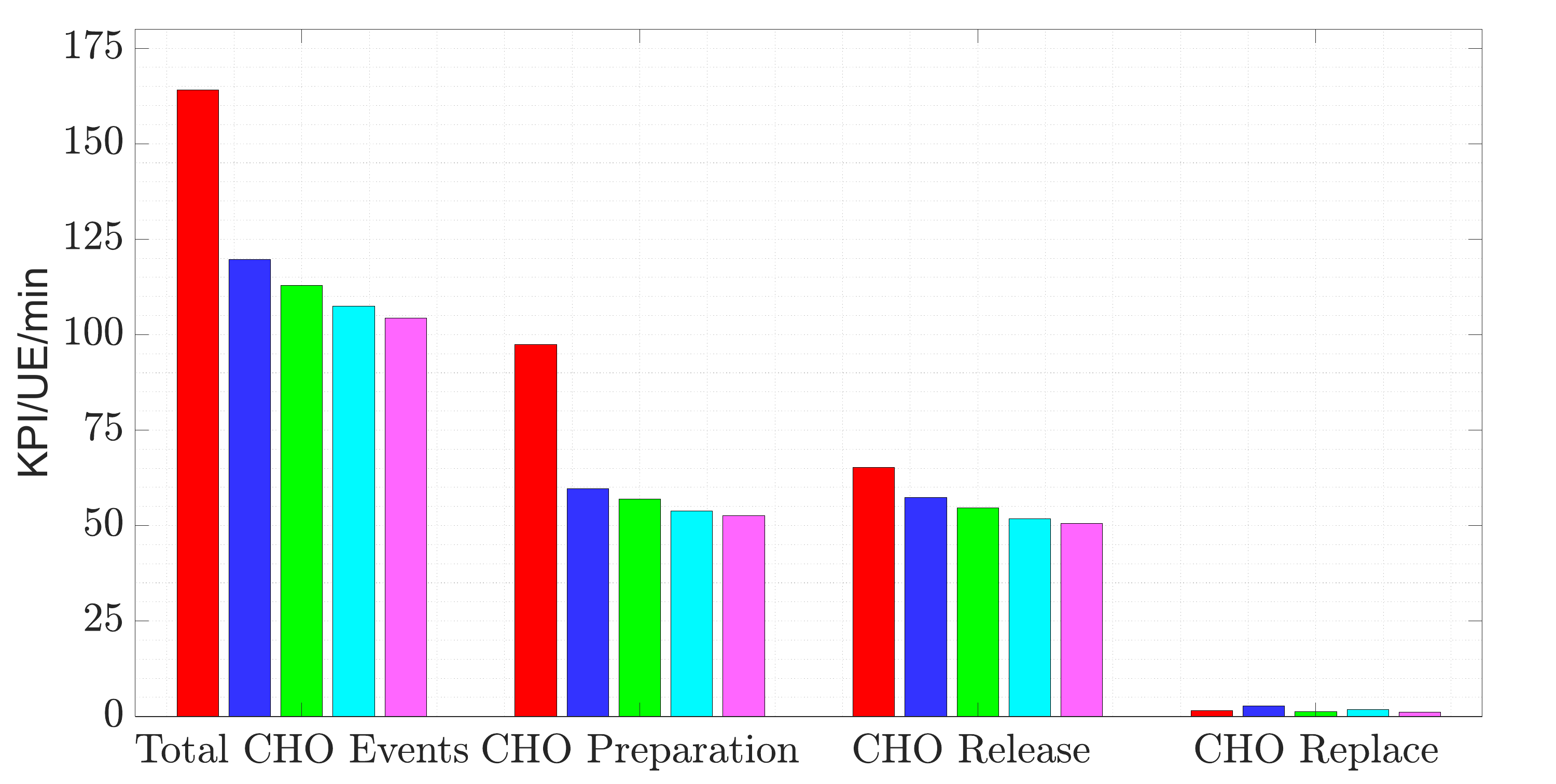}
\label{fig:Fig9b}
\vspace{-14pt}
\caption{Signaling overhead.}  
\label{fig:Fig9b}     
\end{subfigure}\hfill 

\begin{subfigure}{0.5\textwidth}        
\centering        
\includegraphics[width=1\textwidth]{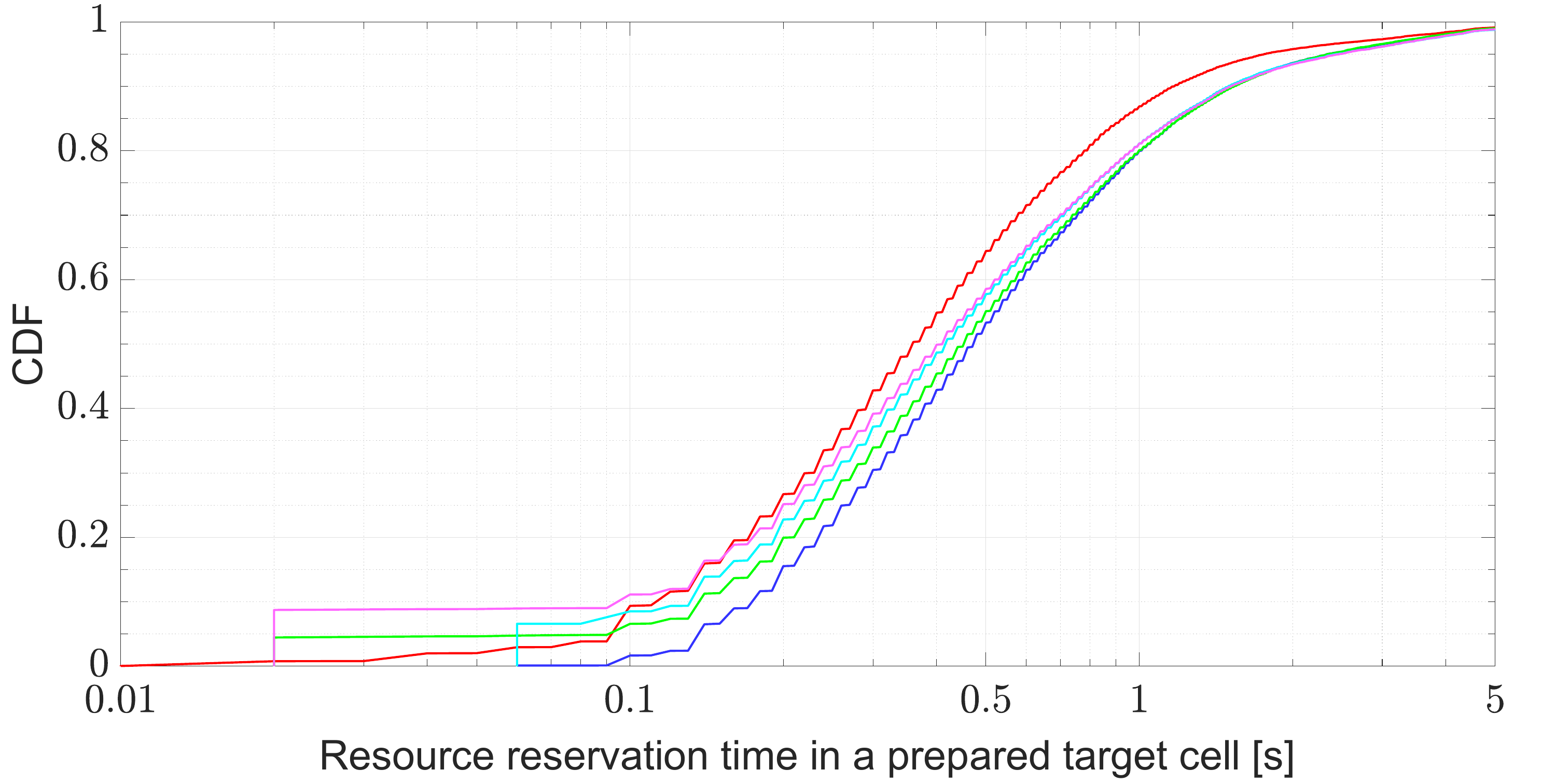} 
\label{fig:Fig9c}
\vspace{-14pt}
\caption{Resource reservation time CDF.} 
\label{fig:Fig9c}     
\end{subfigure}\hfill

\begin{subfigure}{0.5\textwidth}        
\centering        
\includegraphics[width=1\textwidth]{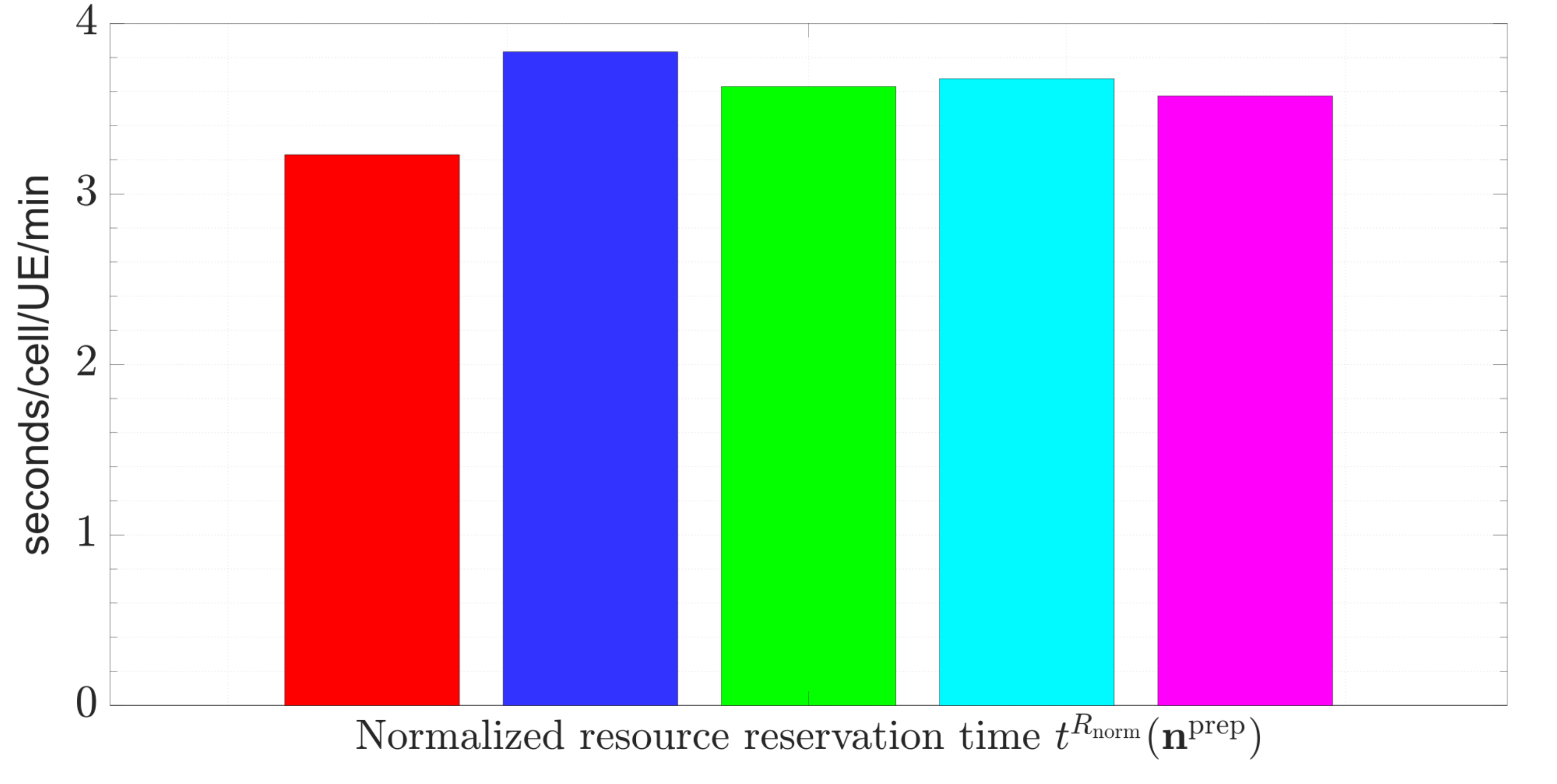} 
\label{fig:Fig9d}
\vspace{-14pt}   
\caption{Normalized resource reservation time in a prepared target cell.}    
\label{fig:Fig9d}   
\end{subfigure}\hfill

\begin{subfigure}{0.5\textwidth}     
\centering        
\includegraphics[width=1\textwidth]{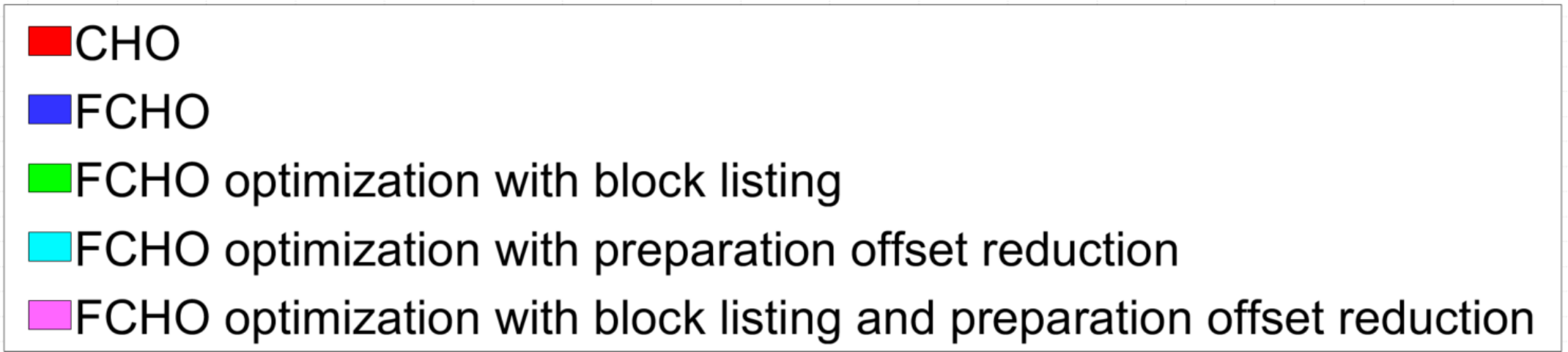}        
\end{subfigure}\hfill  

\caption{The mobility performance of CHO, FCHO, and, the three different FCHO optimization approaches for (a) mobility and outage KPIs, (b) CHO signaling overhead, and (c) and (d) resource reservation time.}
\label{fig:Fig9}   
\end{figure}

Fig.\,\ref{fig:Fig9} shows the performance analysis of the three different FCHO resource reservation optimization approaches in terms of mobility and outage KPIs, CHO signaling overhead, and resource reservation time. Case III of \Cref{Table2} for the highway mobility scenario is considered here.

The first key observation from Fig.\,\ref{fig:Fig9a} is that FCHO optimization with block listing (shown in green) does not lead to any mobility performance degradation when compared to FCHO (shown in blue), i.e., the mobility failures and fast handovers are the same. Consequently, the outage is also~the same. As mentioned in \Cref{Subsec5.2}, this is because the optimization approach block lists the preparation of those cells for which the observed handover probability is zero~and therefore no additional mobility failures are incurred. On the other hand, the two other approaches, namely FCHO optimization with preparation offset reduction (shown~in cyan) and FCHO optimization with block listing and preparation offset reduction (shown in magenta) result in a relative increase of 17.43\% in mobility failures as compared to FCHO. This is because in both these resource reservation optimization approaches some essential target cell preparations for handover are delayed and consequently instead of a successful handover a mobility failure occurs. However, the failures are still 5.23\% less when compared to CHO, meaning the FCHO gain in terms of mobility performance is not totally lost. For these two optimization approaches, fast handovers are reduced as compared to FCHO due to fewer cell preparations in both these approaches. However, since the contribution of a mobility failure to the outage is almost 4 times higher than that of fast handovers, these two optimization approaches result in a 2.3\% greater outage when compared to FCHO.

When the signaling overhead is compared in Fig.\,\ref{fig:Fig9b}, the gain of FCHO (shown in blue) is evident since there is a significant reduction in the total number of CHO events when compared against CHO (shown in red). This relative reduction is 27.04\% for FCHO. FCHO optimization with block listing (shown in green) reduces the total number of signaling overhead even more than FCHO because certain cell preparations are now both \textit{actively} and \textit{reactively} blocked, as discussed in \Cref{Subsec5.2}. This leads to fewer CHO preparation events and consequently fewer CHO release events. The relative reduction in the total number of CHO events for the block listing approach when compared to FCHO is 5.76\%. FCHO optimization with preparation offset reduction (shown in cyan) performs even better than the block listing approach since the preparation offset approach avoids a greater number of preparations as compared to the block listing approach because it takes into account a wider observed handover probability range in the optimization, as discussed in \Cref{Subsec5.2}. The relative reduction in the total number of CHO events for this approach is 10.28\% and 4.80\% when compared to FCHO and FCHO optimization with block listing, respectively. Lastly, it can be seen that FCHO  optimization with block listing and preparation offset reduction approach (shown in magenta) is the best out of the three approaches in terms of signaling overhead reduction since it combines the benefits of both the block listing and offset reduction approaches. The relative reduction in the total number of CHO events for this approach when compared to FCHO is 12.87\%. CHO replace events are relatively less frequent as compared to CHO preparation and CHO release events and have no major bearing on the total number of CHO events. In all of the three optimization approaches, a small reduction in CHO replace events is observed when compared to FCHO since all the optimization approaches reduce preparations and therefore fewer replace events occur because the list of prepared cells is fully occupied less often.

A CDF of the resource reservation time in a prepared target cell $c^{\prime}$ is shown in Fig.\,\ref{fig:Fig9c}. It was already seen in Fig. \ref{fig:Fig8} that at the 50\textsuperscript{th} percentile, the resource reservation time of FCHO (shown in blue) is \SI{0.47}{s} and that of CHO (shown in red) is \SI{0.37}{s}. With the block listing optimization approach (shown in green) it decreases to \SI{0.45}{s} due to the avoidance of many unnecessary temporal resource reservations of the order of \SI{0.50} {s}. At the 95\textsuperscript{th} percentile, the CDFs for FCHO and the block listing optimization approach converge because block listing does not help in \textit{actively} blocking and \textit{reactively} discarding the preparation of those cells that have a high probability of handover and hence they are reserved for longer time periods of the order of \SI{3}{s}. The preparation offset reduction approach (shown in cyan) is more proactive in reducing the resource reservation time because it takes into account a wider observed handover probability range and therefore certain cell preparations either do not occur or last shorter durations due to late cell preparation. Therefore, at the 50\textsuperscript{th} percentile the resource reservation time is \SI{0.42}{s}. It is seen that at the 95\textsuperscript{th} percentile the CDFs for FCHO and the preparation offset reduction approach converge because preparations that last longer and have a high probability of handover cannot be avoided using this optimization approach. Lastly, with the block listing and preparation offset reduction approach (shown in magenta) the resource reservation time reduces to \SI{0.40}{s} at the 50\textsuperscript{th} percentile and is the best out of the three optimization approaches. 

Fig.\,\ref{fig:Fig9d} shows the normalized resource reservation time $t{^{R_{\mathrm{norm}}}} (\mathbf{n}^\mathrm{prep})$ in a prepared target cell. FCHO optimization with block listing (shown in green) reduces the resource reservation time and brings it within the resource reservation time of CHO (shown in red) and FCHO (shown in blue). The relative reduction, when compared with FCHO is 4.97\%. FCHO optimization with preparation offset reduction (shown in cyan) brings about a corresponding gain as the block listing approach. Lastly, it is observed that FCHO optimization with block listing and preparation offset reduction (shown in magenta) is the best out of all the three optimization approaches, where the relative reduction in the resource reservation time/prepared target cell when compared to FCHO is~6.54\%. 

Finally, it can be concluded from Fig.\,\ref{fig:Fig9} that the three different optimization approaches offer a tradeoff between the resource reservation time, mobility performance, and signaling overhead. This is in line with the optimization problem defined in (\ref{Eq8}), where the aim is to optimize the resource reservation time subject to constraints in terms of mobility failures and signaling overhead. FCHO optimization with block listing minimizes the normalized resource reservation time $t{^{R_{\mathrm{norm}}}} (\mathbf{n}^\mathrm{prep})$ while not compromising on the mobility performance gains of FCHO and reduces the signaling overhead below that of FCHO. FCHO optimization with preparation offset reduction offers comparable performance to FCHO optimization with preparation offset reduction but compromises upon the mobility performance gains of FCHO while offering a higher reduction in signaling overhead. FCHO optimization with block listing and preparation offset reduction approach reduces the normalized resource reservation time below that of FCHO optimization with preparation offset reduction and offers comparable mobility performance but a higher reduction in signaling overhead. Hence, for mobility-critical use cases, FCHO optimization with block listing is a suitable approach that offers moderate gains in terms of minimizing both the signaling overhead and resource reservation time. Whereas for other use cases, FCHO optimization with block listing and preparation offset reduction is a suitable approach that offers a relatively higher reduction in both the signaling overhead and resource reservation time, while at the same time, the mobility performance is still not worse off than conventional CHO.

The mobility performance of FCHO optimization with block listing and preparation offset reduction is also shown in Fig.\,\ref{fig:Fig10} for four different values of maximum prepared cells per UE $n_u^{\mathrm{max}}$. It can be observed that with $n_u^{\mathrm{max}}=$ 1 the mobility failures have a significantly high value of 3.86 failures/UE/min on account of having just one prepared target cell to which a handover can be performed. With $n_u^{\mathrm{max}}=$ 2, mobility failures reduce relatively by 19.73\% on account of an additional prepared cell which can serve as a potential handover candidate. This corresponds to a relative increase of 11.55\% in the fast handovers. With $n_u^{\mathrm{max}}=$ 4, the mobility failures reduce by 3.54\% when compared to $n_u^{\mathrm{max}}=$ 2. For the highest 3GPP defined value of $n_u^{\mathrm{max}}=$ 8 \cite{b1}, it is seen that the mobility performance levels off as more prepared cells do not yield a gain in mobility performance in terms of mobility failures and fast handovers and correspondingly the outage.

\begin{figure}[!b]
\textit{\centering
\includegraphics[width = 1\columnwidth]{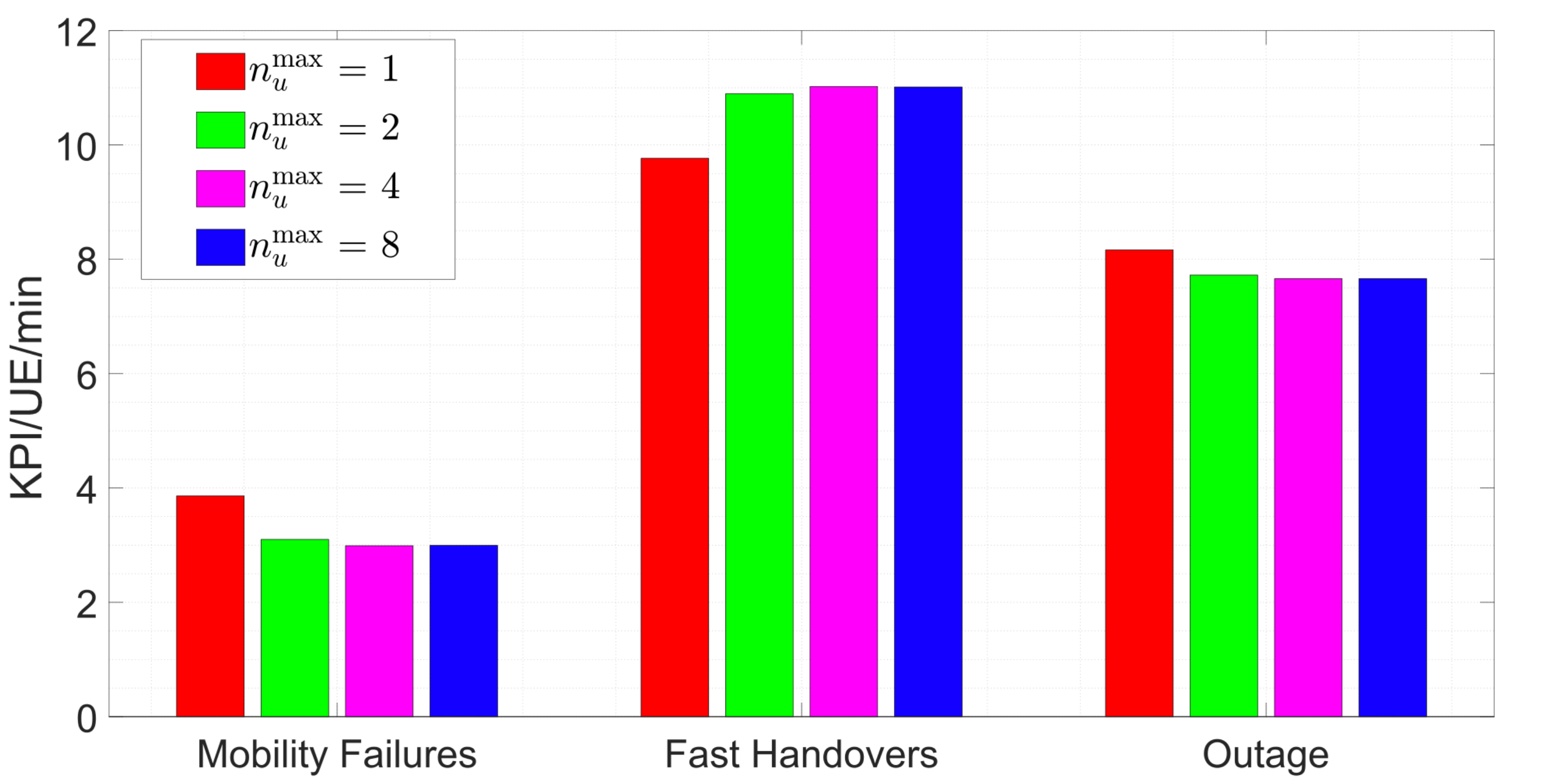}
\label{fig:Fig10}
\vspace{-\baselineskip}
\caption{The mobility performance of FCHO optimization with block listing and preparation offset reduction for different values of maximum prepared cells $n_u^{\mathrm{max}}$ per UE.} 
\label{fig:Fig10}} 
\end{figure}

\section{Conclusion} \label{Sec7}

In this article, the mobility performance for MPUE with three panels is investigated with hand blockage using CTIA wide-grip hand phantom model for two different mobility scenarios in a 5G-Advanced network. Four different hand positions with varying degrees of blockage on each of the panels are considered and a detailed analysis is carried out for the CHO and FCHO mechanisms. It is seen that FCHO reduces mobility failures by 10.5\% and 19.3\% compared to CHO for the urban and highway mobility scenarios, respectively, for the worst-case scenario where two out of three panels are completely blocked. As a future topic open for research, additional studies that consider the probability of hand blockage for each panel are required to bring the performance analysis closer to real-world scenarios. Furthermore, the problem of longer resource reservation time in FCHO, caused by keeping the conditional configuration after each cell change, is discussed in detail. To tackle this issue, an optimization problem is formulated and three different FCHO resource reservation optimization techniques following MRO principles are introduced. Then a detailed performance analysis is carried out for the highway scenario and for the case when two panels out of three are blocked. It is observed that each optimization technique is beneficial to varying degrees in reducing both the signaling overhead and resource reservation time and offers a unique tradeoff between mobility and outage KPIs, CHO signaling overhead, and resource reservation. For example, with FCHO optimization with block listing and preparation offset reduction, the mobility performance is still better than CHO but the signaling overhead and resource reservation time reduce by 12.87\% and 6.54\%, respectively, when compared with FCHO. This implies that different resource reservation optimization techniques can be adopted to suit the needs of~the network.

\section{Acknowledgment} \label{Sec8}
The authors would like to express their gratitude to Simon Svendsen from Nokia Standardization and Research Lab, Aalborg, Denmark for providing us with the  \textit{CST Studio Suite} generated hand blockage radiation patterns.

\begin{IEEEbiography}[{\includegraphics[width=1in,height=1.25in,clip,keepaspectratio]{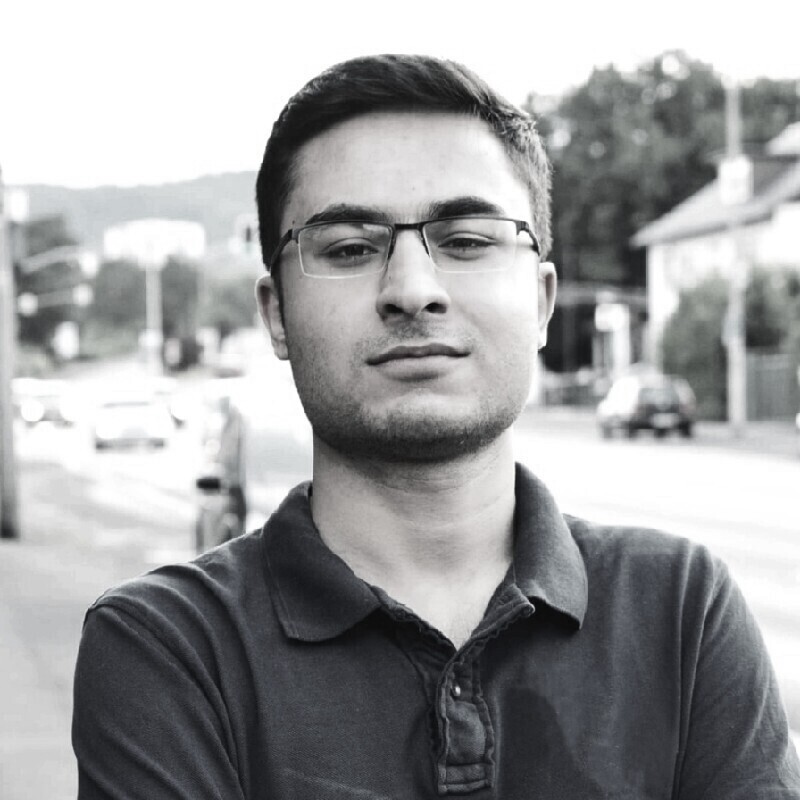}}]{Subhyal Bin Iqbal} (Graduate Student Member, IEEE) received the M.Sc. degree in Electrical Communication Engineering from the University of Kassel, Germany, in 2020. During this time he also interned at Bosch Research, Renningen, where he was working on compressed sensing and distributed source coding for massive wireless sensor networks. In 2020 he joined Nokia Bell Labs, Munich as a Ph.D. candidate in cooperation with Vodafone Chair for Mobile Communications Systems at Technische Universität Dresden. His research interests include mobility robustness for wireless communication systems, UE architecture design, and radio resource management.
\end{IEEEbiography}

\begin{IEEEbiography}
[{\includegraphics[width=1in,height=1.25in,clip,keepaspectratio]{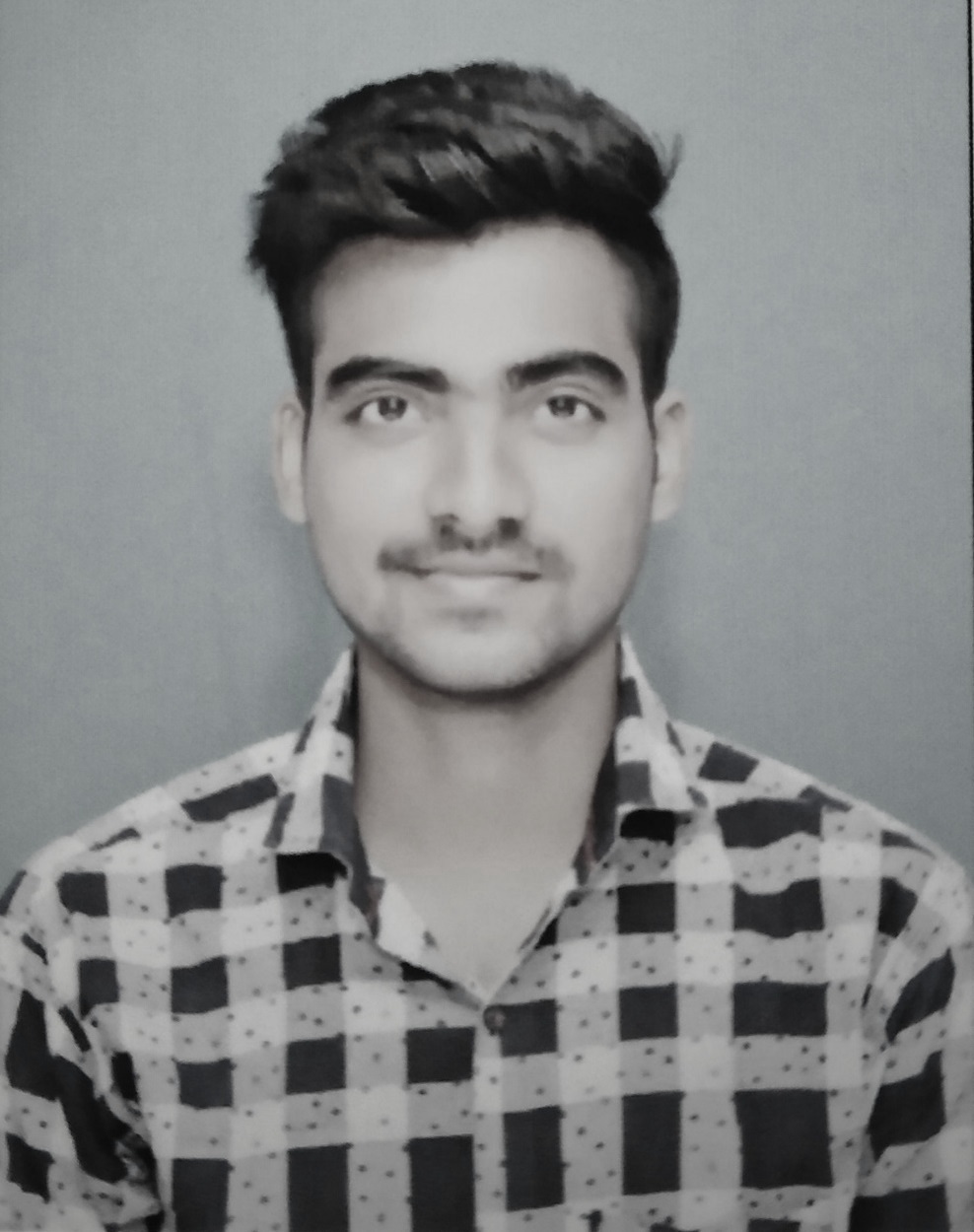}}]{Salman Nadaf} received the B.E. degree in Electrical Engineering from Mumbai University, India, in 2018, and he is currently pursuing M.Sc. degree in Communication Engineering at the Technical University of Munich, Germany. In 2022, he joined the Radio Access and Architecture Munich Department, Standardization Research Laboratory, where he is working on a Master's thesis. His research interests include mobility performance investigations in mmWave communication, security, resilience, and control of IoT and cyber-physical systems. 
\end{IEEEbiography}

\begin{IEEEbiography}
 [{\includegraphics[width=1in,height=1.25in,clip,keepaspectratio]{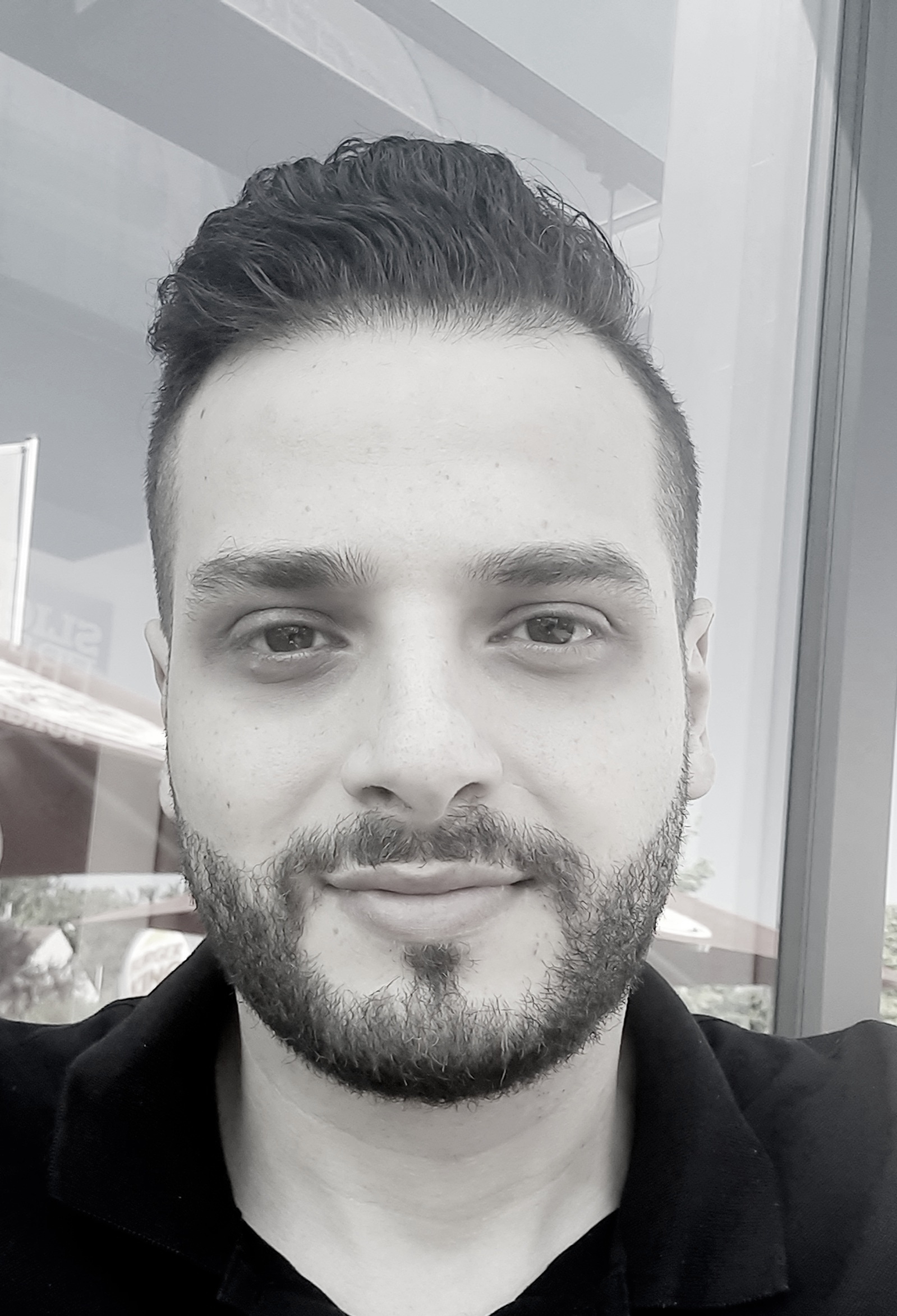}}]{Ahmad Awada} (Member, IEEE) received the M.Sc. degree in Communication Engineering from the Technical University of Munich, in 2009, and the Ph.D. degree from the Technical University of Darmstadt, Germany, in 2014. He joined Nokia Networks in 2013. Since 2016, he has been working for the Radio Access and Architecture Munich Department, Standardization Research Laboratory, dealing with LTE and 5G standardization research. His research interests include radio transmission schemes, radio resource management and control, and network slicing.   
\end{IEEEbiography}

\begin{IEEEbiography}
  [{\includegraphics[width=1in,height=1.25in,clip,keepaspectratio]{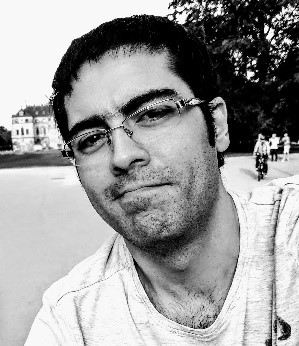}}] {Umur Karabulut} (Member, IEEE) received the M.Sc. degree in Communications Engineering from the Technical University of Munich, Germany, in 2017. In 2017 he joined Nokia Bell Labs, Munich as a Ph.D. student in cooperation with Vodafone Chair for Mobile Communications Systems at Technische Universität Dresden. Since 2020, he is working as a Radio Access Specialist at Nokia in the Radio Access and Architecture Munich Department, Standardization Research Laboratory. His current research interest is mobility enhancements beyond 5G systems.  
\end{IEEEbiography}

\vspace{-120pt}

\begin{IEEEbiography}
  [{\includegraphics[width=1in,height=1.25in,clip,keepaspectratio]{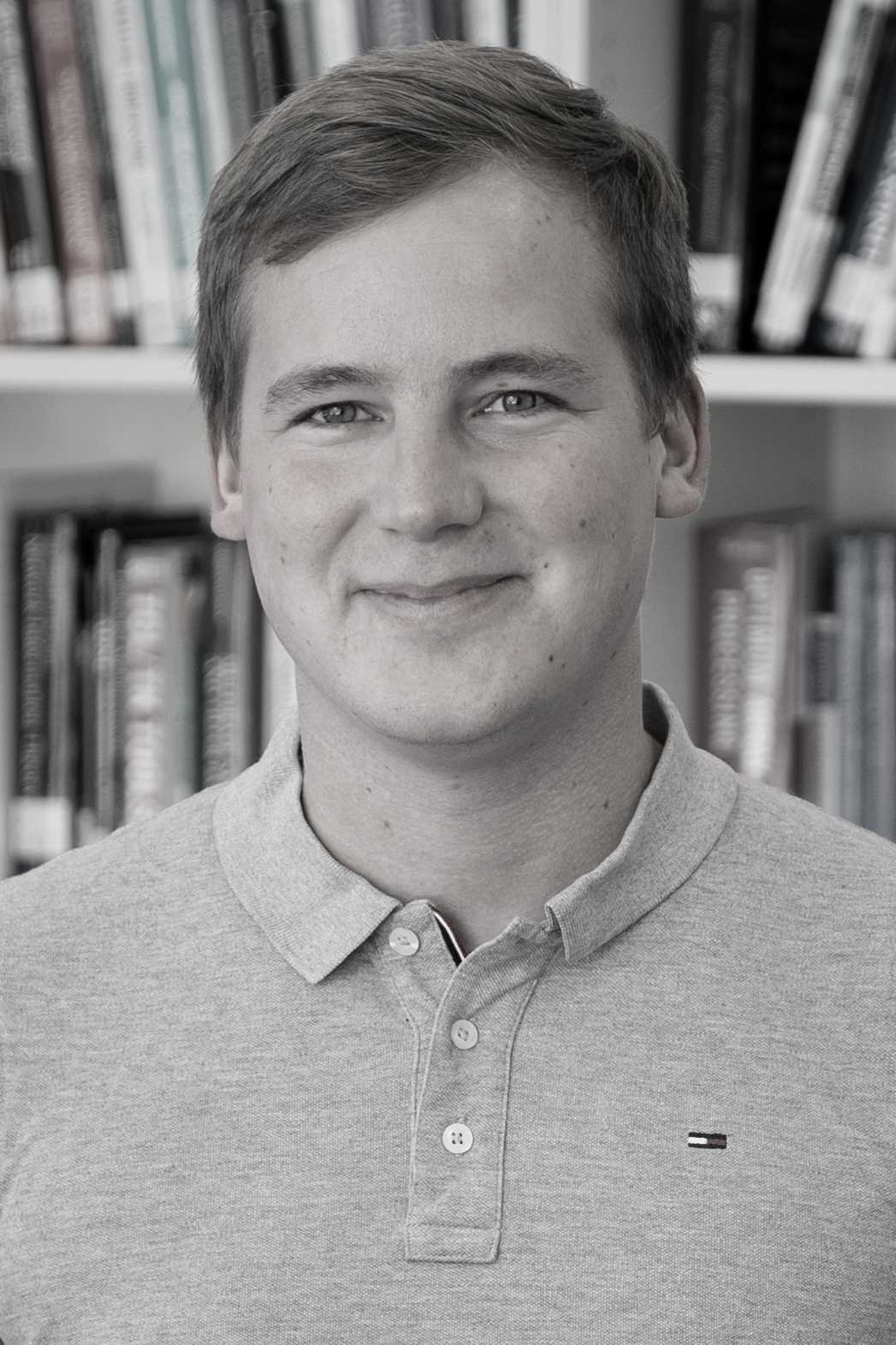}}] {Philipp Schulz} (Member, IEEE) received the M.Sc. degree in Mathematics and the Ph.D. (Dr.-Ing.) degree in Electrical Engineering from Technische Universität Dresden, Germany, in 2014 and 2020, respectively. He was a Research Assistant with Technische Universität Dresden in the field of numerical mathematics, modeling, and simulation, where he joined the Vodafone Chair for Mobile Communications Systems in 2015 and became a member of the System-Level Group. His research there focused on flow-level modeling and the application of queuing theory on communications systems with respect to ultra-reliable low-latency communications. After more than one year at the Barkhausen Institut, Dresden, Germany, where he studied rateless codes in the context of multi-connectivity, he is currently a research group leader at the Vodafone Chair and focuses on the resilience of wireless communications systems.
\end{IEEEbiography}

\vspace{-120pt}

\begin{IEEEbiography}
  [{\includegraphics[width=1in,height=1.25in,clip,keepaspectratio]{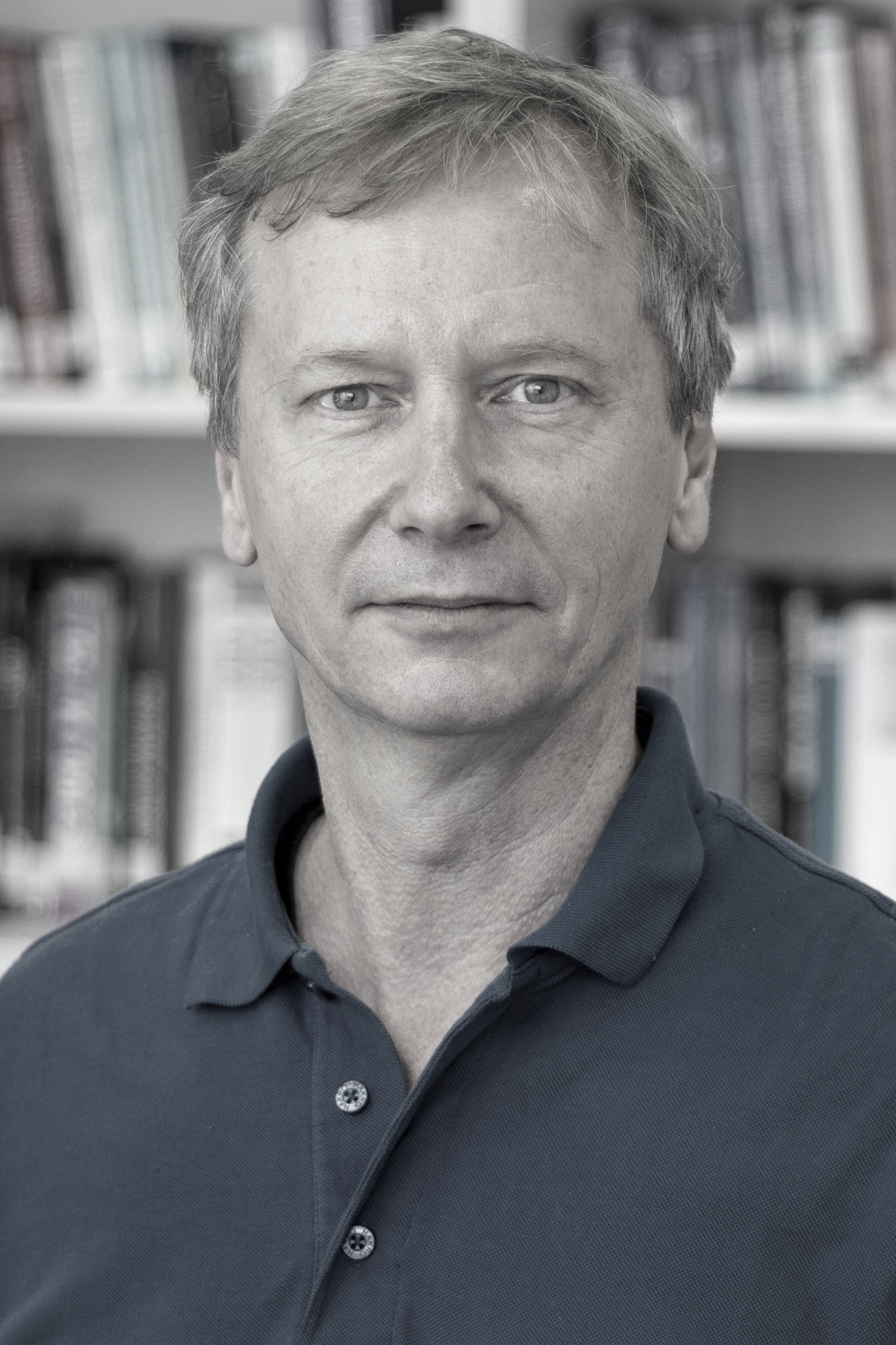}}] {Gerhard Fettweis} (Fellow, IEEE) earned a Ph.D. under H. Meyr at RWTH Aachen in 1990. After postdoctoral work at IBM Research, San Jose, California, he joined TCSI, Berkeley, United States. Since 1994, he has been Vodafone Chair Professor at Technische Universität Dresden. Since 2018 he has also headed the new Barkhausen Institute. In 2019 he was elected into the DFG Senate (German Research Foundation). He researches wireless transmission and chip design, coordinates 5G++Lab Germany, has spun out 17 tech and 3 non-tech startups, and is a member of the German Academy of Sciences (Leopoldina), and German Academy of Engineering (acatech).
\end{IEEEbiography}

\EOD

\end{document}